\documentclass[twocolumn]{aastex631}

\begin{document}

\title{BASS XLVII: 22 GHz Radio Atlas of Swift-BAT Selected AGN}

\author[0000-0002-1292-1451]{Macon Magno}
\affiliation{George P. and Cynthia Woods Mitchell Institute for Fundamental Physics and Astronomy, Texas A\&M University, College Station, TX, 77845, USA}
\affiliation{CSIRO Space and Astronomy, ATNF, PO Box 1130, Bentley WA 6102, Australia}
\affiliation{Southern Methodist University, Department of Physics, Dallas, TX, 75205, USA}

\author[0000-0001-5785-7038]{Krista L. Smith}
\affiliation{George P. and Cynthia Woods Mitchell Institute for Fundamental Physics and Astronomy, Texas A\&M University, College Station, TX, 77845, USA}
\affiliation{Southern Methodist University, Department of Physics, Dallas, TX, 75205, USA} 

\author[0000-0003-4264-3509]{O. Ivy Wong}
\affiliation{CSIRO Space and Astronomy, ATNF, PO Box 1130, Bentley WA 6102, Australia}
\affiliation{ICRAR-M468, The University of Western Australia, 35 Stirling Hwy, Crawley, WA 6009, Australia}

\author[0000-0002-7962-5446]{Richard Mushotzky}
\affiliation{Department of Astronomy, University of Maryland, College Park, MD 20742, USA}
\affiliation{Joint Space-Science Institute, University of Maryland, College Park, MD 20742, USA}

\author[0000-0002-8765-7915]{Stuart Vogel}
\affiliation{Department of Astronomy, University of Maryland, College Park, MD 20742, USA}

\author[0000-0002-7998-9581]{Michael J. Koss}
\affiliation{Eureka Scientific, 2452 Delmer Street, Suite 100, Oakland, CA 94602-3017, USA}
\affiliation{Space Science Institute, 4750 Walnut Street, Suite 205, Boulder, CO 80301, USA}

\author[0000-0001-5231-2645]{Claudio Ricci}
\affiliation{Instituto de Estudios Astrof\'isicos, Facultad de Ingenier\'ia y Ciencias, Universidad Diego Portales, Av. Ej\'ercito Libertador 441, Santiago, Chile}
\affiliation{Kavli Institute for Astronomy and Astrophysics, Peking University, Beijing 100871, China}

\author[0000-0002-5037-951X]{Kyuseok Oh}
\affiliation{Korea Astronomy \& Space Science institute, 776, Daedeokdae-ro, Yuseong-gu, Daejeon 34055, Republic of Korea}

\author[0000-0001-9910-3234]{Chin-Shin Chang}
\affiliation{Joint ALMA Observatory, Avenida Alonso de Cordova 3107, Vitacura 7630355, Santiago, Chile}

\author[0000-0003-0057-8892]{Loreto Barcos-Mu\~noz}
\affiliation{National Radio Astronomy Observatory, 520 Edgemont Road, Charlottesville, VA 22903, USA}
\affiliation{Department of Astronomy, University of Virginia, 530 McCormick Road, Charlottesville, VA 22903, USA}

\author[0000-0002-8686-8737]{Franz E. Bauer}
\affiliation{Instituto de Astrof\'isica, Facultad de F\'isica, Pontificia Universidad Cat\'olica de Chile, Campus San Joaquin, Av. Vicu\~na Mackenna 4860, Macul Santiago,
7820436, Chile}
\affiliation{Centro de Astroingenier\'ia, Facultad de F\'isica, Pontificia Universidad Cat\'olica de Chile, Campus San Joaquin, Av. Vicu\~na Mackenna 4860, Macul Santiago,
7820436, Chile}
\affiliation{Millennium Institute of Astrophysics, Nuncio Monse\~nor S\'otero Sanz 100, Of 104, Providencia, Santiago, Chile}

\author[0000-0003-2196-3298]{Alessandro Peca}
\affiliation{Eureka Scientific, 2452 Delmer Street, Suite 100, Oakland, CA 94602-3017, USA}
\affiliation{Department of Physics, Yale University, P.O. Box 208120, New Haven, CT 06520, USA}

\author[0000-0002-2603-2639]{Darshan Kakkad}
\affiliation{Space Telescope Science Institute, 3700 San Martin Drive, Baltimore, 21218 MD, USA}

\author[0000-0002-9144-2255]{Turgay Caglar}
\affiliation{George P. and Cynthia Woods Mitchell Institute for Fundamental Physics and Astronomy, Texas A\&M University, College Station, TX, 77845, USA}
\affiliation{Leiden Observatory, Leiden University, PO Box 9513, NL-2300 RA Leiden, the Netherlands} 

\author[0000-0002-3683-7297]{Benny Trakhtenbrot}
\affiliation{School of Physics and Astronomy, Tel Aviv University, Tel Aviv 69978, Israel}

\author[0000-0002-4226-8959]{Fiona Harrison}
\affiliation{Cahill Center for Astronomy and Astrophysics, California Institute of Technology, Pasadena, CA 91125, USA}

\author[0000-0003-2686-9241]{Daniel Stern}
\affiliation{Jet Propulsion Laboratory, California Institute of Technology, 4800 Oak Grove Drive, MS 169-224, Pasadena, CA 91109, USA}

\author[0000-0002-0745-9792]{C. Megan Urry}
\affiliation{Yale Center for Astronomy \& Astrophysics, Physics Department, PO Box 208120, New Haven, CT 06520-8120, USA}

\author[0000-0003-2284-8603]{Merry Powell}
\affiliation{Kavli Institute for Particle Astrophysics and Cosmology, Stanford University, 452 Lomita Mall, Stanford, CA 94305, USA}
\affiliation{Department of Physics, Stanford University, 382 Via Pueblo Mall, Stanford, CA 94305, USA}

\correspondingauthor{Macon Magno}
\email{macon.a.magno@tamu.edu}



\begin{abstract}
 
We present the third phase of the largest high-frequency, high-resolution imaging survey of 231 nearby, hard X-ray selected AGN, with a very high $98 \pm 1\%$ detection fraction. This survey presents VLA 22~GHz radio observations with 1\arcsec~spatial resolution covering over $6$ orders of magnitude in radio luminosity in nearby AGN that span $\sim4$ orders of magnitude in black hole mass and X-ray luminosity.  We identify three different radio morphologies: $44 \pm 3\%$ (102/231) are compact or unresolved, $46 \pm 3\%$ (106/231) show an extended structure (star formation, possible one-sided jets, etc.), and $8 \pm 2\%$ (19/231) have a biconical or two-sided jet-like morphology. The remaining $2 \pm 1\%$ (4/231) sources are non-detections. The radio-to-X-ray luminosity ratios of the Swift-BAT AGN in our sample ($\text{L}_R/\text{L}_{14-195 \text{keV}} \sim 10^{-5.5}$ and $\text{L}_R/\text{L}_{2-10 \text{keV}} \sim 10^{-5}$ with a scatter of $\sim0.5$ dex) are similar to that of coronally active stars ($\text{L}_R/\text{L}_X \sim 10^{-5}$). For most targets, extended emission in radio-quiet objects is broadly consistent with the expectation for star formation from previous FIR observations, once the contribution from the radio core has been subtracted. Our sample represents nearby analogs of distant AGN at the peak of black hole growth, and thus the high detection fraction in our work has important implications for future high frequency AGN radio surveys with the next generation VLA (ngVLA) or Square Kilometre Array (SKA), both of which should detect large fractions of more distant AGN.

\end{abstract}



\section{Introduction} \label{sec:intro}

Supermassive black holes (SMBHs) are found in the centers of massive galaxies, and, in some cases, they are found to be accreting a substantial amount of matter; in this case known as an Active Galactic Nucleus (AGN). These AGN can emit radiation ranging from radio waves to gamma rays \citep[e.g.,][]{2013MNRAS.431.2471B, 2015MNRAS.447.1289P, 2024MNRAS.529.1365P}. The origin of the emission in each waveband is physically distinct; the accretion disk radiates primarily in the UV / optical wavelength, whereas the radio and X-ray emission originates primarily from a small-scale corona or a radio jet \citep[e.g.,][\& references therein]{1999agnc.book.....K, 2013peag.book.....N}. While the radio emission of Radio-loud AGN is dominated by a powerful relativistic jets which can extend out to megaparsec scales, \citep[e.g.,][]{1995A&A...293..665F, 1982MNRAS.199..883B, 1991ApJ...383..554C, 1995MNRAS.277.1097L, 2000ApJ...540..678B, 2022Galax..10....6F} the vast majority of AGN are ``radio-quiet''. Observations suggest that such radio-loud AGN only make up $\sim10$\% of all AGN at low redshift \citep[e.g.,][]{2005MNRAS.362...25B, 2019A&A...622A..10C, 2020NewAR..8801539H, 2024MNRAS.528.5432T}. Radio-quiet AGN have radio emission that does not dominate their luminosity or morphology, and radio jets are not necessarily clearly present \citep[e.g.,][]{2016ApJ...832..163S, 10.1093/mnras/stz3608}. For radio-quiet AGN, the dominant emission source in radio and X-ray emission is possibly the central corona, although other ideas proliferate \citep[e.g.,][]{2022MNRAS.510.1043B, 2023ApJ...952L..28R, 2023ApJ...957...69Z}. 

Large-scale jets have long been identified as probable culprits for AGN feedback: the process by which the AGN regulates star formation in its host. However, given their relative rarity, they cannot be the primary driver in most galaxies. Instead, among the less luminous population of Seyfert galaxies, radio emission on kiloparsec scales is now recognized to be quite common \citep{1983HiA.....6..467W, 2021MNRAS.503.1780J, 2021A&A...650A..84V}. In addition to smaller or weaker radio jets, there is extended emission, likely due to star formation and radiative outflows originating from the AGN, which are likely to be more important feedback drivers \citep[e.g.,][]{2015A&A...582A..63C,2020ApJ...904...83S, 2022ApJ...936...76S, 2023Galax..11...27K, 2024Galax..12...17H}. 

There are two main ideas on the relationship between radio-quiet and radio-loud AGN. One idea is that radio-quiet AGN are the smaller, quieter ``relatives'' of radio-loud AGN, such that all radio-quiet AGN will eventually develop into a radio-loud AGN. In such a scenario, radio-quiet AGN have smaller, underdeveloped radio jets that expand through their host galaxy and eventually will reach the scales of a radio-loud AGN \citep{2003MNRAS.343L..59H, 2005Ap&SS.300...15H}. This idea is supported by the ``Fundamental Plane of Black Hole Activity'', which takes into account the tight correlation between radio emission, X-ray emission, and black hole mass \citep{2003MNRAS.345.1057M, 2004A&A...414..895F}. The fundamental plane is based on a theoretical scale-invariant disk-jet connection in accreting black holes, where the X-ray emission is related to the accretion disk, and the radio emission arises from the radio jet, even when nascent. It should be noted that the ``Fundamental Plane of Black Hole Activity'' only applies to strong radio-loud AGN while weaker radio sources deviate from this by several dex \citep{2022MNRAS.513.4673B}. The other idea is that radio emission in radio-quiet AGN and radio-loud AGN comes from two different sources. In the case of radio-loud AGN, the radio emission is dominated by synchrotron emission from the radio jet, meanwhile, in radio-quiet AGN, the AGN's corona is thought to be the major contributor to the radio emission \citep{2008MNRAS.390..847L, 2016MNRAS.459.2082R}.

The corona may be the launching point of the radio jet \citep[e.g.,][]{2015MNRAS.449..129W}, but this is not necessarily an established fact. The corona is believed to be the origin of X-ray emission, and depending on the magnetic field strength, orientation, and geometry, radio emission is expected anywhere between 10 - 100 GHz, and, in fact, it has been shown that there is a tight correlation between 100 GHz emission and X-rays \citep[e.g.,][]{2014PASJ...66L...8I, 2016MNRAS.459.2082R, 2018MNRAS.478..399B, 2023ApJ...952L..28R}. It is also possible that star formation, radiatively driven AGN winds, and/or AGN outflow shocks with the interstellar medium can also be a potential source of radio emission \citep[e.g.,][]{2016MNRAS.455.4191Z, 2022ApJ...938...87K, 2023arXiv231210177C, 2023ApJS..269...24K}.

To probe the prevalence of kiloparsec-scale radio structures in radio-quiet AGN and measure the regime most pertinent to coronal emission simultaneously, we conducted a 22~GHz radio survey with 1\arcsec~angular resolution of a nearby ($z<0.05$) sample of 231 predominantly radio-quiet AGN and 7 radio-loud AGN from the \textit{Swift}-BAT 70-month survey. Our results include a very high $98 \pm 1\%$ (227/231) detection fraction, which includes the results of the previous two phases published in \cite{2016ApJ...832..163S} and \cite{10.1093/mnras/stz3608}, hereafter Papers I and II, respectively. Since the sample size has roughly doubled from Paper II, and includes updated physical parameter estimates from subsequent data releases of the BAT AGN Spectroscopic Survey \citep[BASS;][]{Koss_2017, Ricci_2017, Koss_2022}, our analysis is significantly more statistically robust than the first two phases of this survey and includes several objects with radio luminosities higher than $10^{39}$ erg $\text{s}^{-1}$, which were not targeted in Papers I and II. Our intent is to provide a public atlas of high-frequency (22~GHz) radio morphologies and core radio flux densities of a \textit{Swift}-BAT ultra-hard X-ray selected AGN sample. The survey's ultra-hard X-ray selection yields a relatively unbiased sample that is independent of most properties of the host galaxy such as mass, extinction, luminosity, star formation rate, etc. 

In Section \ref{sec:sample}, we describe the sample and 22~GHz imaging campaign, along with the supplementary data from BASS. Section \ref{sec:morphologies} describes the 22~GHz radio morphologies. In Section \ref{sec:Results}, we explore various correlations between the radio, X-ray, and FIR data, along with a discussion of our detection fraction and non-detections. We list our conclusions in Section \ref{sec:Conclusion}. 

Throughout this paper, we assume a cosmology of $H_0 = 69.6$ km s$^{-1}$ Mpc$^{-1}$, $\Omega_M$ = 0.286, and $\Omega_\Lambda = 0.714$. For all calculations involving redshift, we used \cite{2006PASP..118.1711W}'s and \cite{astropy:2013, astropy:2018, astropy:2022}'s Cosmology Calculators.

\section{Sample Selection and Data Reduction} \label{sec:sample}

\subsection{The BAT AGN Spectroscopic Survey} \label{subsec:bass}

The BAT AGN Spectroscopic Survey \citep[BASS;][]{Koss_2017, Ricci_2017, Koss_2022}\footnote{\url{www.bass-survey.com}} is a collection of multi-wavelength spectra and images of the \textit{Swift}-BAT AGN sample. The \textit{Swift}-BAT all-sky survey provides a relatively unbiased AGN sample, as the ultra-hard X-ray band (14-195 keV) can detect AGN even up to the Compton-thick regime ($10^{24}$ $\text{cm}^{-2}$) and above \citep[e.g.,][]{2009ApJ...696..891H, 2016ApJ...825...85K, 2017ApJS..233...17R, 2018ARA&A..56..625H, 2018MNRAS.480.1819R}. This selection criterion probes a wide range of black hole masses and accretion rates. \citet{Koss_2022} provides the black hole mass, spectral type, and the Eddington ratio ($\lambda_\mathrm{Edd}$) estimates for the 70-month \textit{Swift}-BAT AGN. Additionally, BASS provides redshifts of the BAT AGN measured from emission lines along with the intrinsic absorption-corrected X-ray fluxes from the \textit{Swift}-BAT and various soft X-ray telescopes \citep{2015ApJ...815L..13R, Ricci_2017}. Table \ref{tab:22GHz_props} lists the BASS properties of the 22 GHz sample.

\subsection{Sample} \label{subsec:sample}

Our parent sample is derived from \cite{Mushotzky_2014}, which consists of 313 low redshift ($z<0.05$) AGN from the 58-month \textit{Swift}-BAT ultra-hard X-ray (14-195 keV) survey \citep{2010HEAD...11.1305B} and were imaged with the \textit{Herschel} PACS (70 and 160 $\mu$m) and SPIRE (250, 350, and 500 $\mu$m) \citep{2017MNRAS.466.3161S}. We have completed the third phase of a K-band (22 GHz), 1\arcsec~resolution radio imaging survey with the Karl G. Jansky Very Large Array (VLA)\footnote{The VLA is operated by the National Radio Astronomy Observatory (NRAO), which is a facility of the National Science Foundation operated under cooperative agreement by the Associated Universities, Inc.} in C-array of 231 \textit{Swift}-BAT ultra-hard X-ray selected AGN. The rest of \cite{Mushotzky_2014}'s sample was not visible from the VLA, thus this completes their original sample. In general, the sample includes 66\% (226/342) of the 70-month BAT AGN from the BASS DR2 above $-31$ degrees declination at $z<0.05$ in the sky. This excludes 4 featureless blazars in this redshift range.

Paper I presented the initial results of this survey, in which they observed their sample of 70 low redshift ($z<0.05$) targets from \cite{Mushotzky_2014}. Paper II included 30 new AGN with the same selection criteria as in the previous stage, expanding the total sample size to 100 AGN.  Objects in the first two phases of the survey are relatively nearby ($z < 0.05$), with 22 GHz radio luminosities in the range of $10^{36.5} - 10^{41}$ erg $\text{s}^{-1}$.

Here we present a 22\,GHz VLA survey of 131 new targets from the parent sample described in \cite{Mushotzky_2014}. Within this expanded sample, we have added a few objects with radio luminosities greater than $10^{39}$ erg $\text{s}^{-1}$. Mrk\,595 and 2MASX\,J01073963-1139117, non-detections in the first phase, were recovered from their original observations during this third phase through careful flagging during re-reduction. In addition to these objects becoming detections, five objects previously discarded due to RFI effects were re-reduced and recovered. Two new objects from this phase are non-detections: IRAS\,03219+4031 and MCG-02-02-095. We now have a total of four non-detections for our entire survey. For this sample, we used objects from the 58-month and the 70-month \textit{Swift}-BAT all-sky survey \citep{2010HEAD...11.1305B, 2013ApJS..207...19B}. Two objects have dropped out of the 70-month \textit{Swift}-BAT all-sky survey, likely due to X-ray variability: Mrk\,202 and 2MASX\,J12335145-2103448. Since our VLA observations were not simultaneous with the \textit{Swift}-BAT all-sky survey, the non-detected AGN could have fallen below our sensitivity of $16$ $\mu$Jy due to rapid variability in the radio. We note that the non-detections were not detected at a $5\sigma$ level ($\sim80$ $\mu$Jy), but Mrk 653 and Mrk 352 had a marginal unresolved detection at $3\sigma$ ($\sim48$ $\mu$Jy) with a flux density of $\sim$72 $\mu$Jy. For the purposes of this paper, we have decided to exclude the marginal detection of Mrk 653 and Mrk 352 from our analysis, and instead use upper limits for all four of our non-detections. 

This phase, in combination with the previous phases, completes the original sample observed by \cite{Mushotzky_2014} with \textit{Herschel} and were observable with the VLA. Unlike the previous two phases, which focused primarily on radio-quiet AGN, this phase did not have a radio loudness criterion. Despite this, our sample is still remarkably radio-quiet and only has six radio-loud AGN: Arp\,102B, Mrk\,348, NGC\,1052 (2 observations), [HB89]\,0241+622, PKS\,2331-240, and 2MASX\,J23272195+1524375.
The luminosity range of this work is similar to that of Papers I and II, ranging from $10^{36} - 10^{42}$ erg $\text{s}^{-1}$. Moreover, this sample spans $\sim4$ orders of magnitude in black hole mass ($10^5 - 10^9$), X-ray luminosity ($10^{40.5} - 10^{44.5}$), and column density ($\text{N}_{\text{H}}$; $10^{20}-10^{24}$).

Table \ref{tab:22GHz_props} shows the radio properties of the sample including the $1\arcsec$, $3\arcsec$, and $6\arcsec$ fluxes (see Section~\ref{subsec:radio}), $1\sigma$ errors, and radio morphology class for the entire sample; the machine-readable table online includes all 231 sources. Additionally, all images presented in this table are publicly available on GitHub\footnote{\url{https://github.com/maconmagno/22GHz/tree/main/AGN}}. The data is also available on Zenodo under an open-source Creative Commons Attribution license: \dataset[doi:10.5281/zenodo.13988276]{https://doi.org/10.5281/zenodo.13988276}. A subset of our sample was also observed at a higher resolution (0.3\arcsec) with the VLA in B-array at 22 GHz in Paper I. For practicality, Table \ref{tab:22GHz-barray} lists the integrated fluxes obtained from maps using only B-array images.

\subsection{Repeated Observations} \label{subsec:repeat}

Fifteen sources were observed more than once in our campaign. Of these, four sources varied more than 3-sigma between the two observations, the most spectacular case being NGC\,1052, which had a $30 \pm 0.28\%$ flux density variation between 2014 and 2017. We can therefore say that the source of the 22 GHz emission is reasonably compact, such that it can vary significantly on $\sim$year timescales. NGC\,1052 is a known radio-loud source with a double-sided jet, so it is not surprising that it varies \citep{2003A&A...401..113V}. Another interesting source is 2MASX\,J04440903+2813003, which is known to vary rapidly in the X-rays \citep{2014A&A...563A..57S}.  These sources have an additional entry in the Table \ref{tab:22GHz_props}, and are represented with the year of the repeated observation in parentheses below the original observation.

\begin{splitdeluxetable*}{ccccccccccBcccccccc}
\tablecaption{BASS Properties of the 22 GHz Sample and the VLA C-array Flux and Morphology Measurements\label{tab:22GHz_props}}
\tablehead{\colhead{BAT ID} & \colhead{Name} & \colhead{RA} & \colhead{Dec} & \colhead{Redshift} & \colhead{Seyfert} & \colhead{log Mass} & \colhead{log $\lambda_\mathrm{Edd}$} &  \colhead{F$_{HX,int}$} & \colhead{F$_{UHX,int}$} & B$_{\text{maj}}$$\times$B$_{\text{min}}$, PA & \colhead{1\arcsec} & \colhead{S$_{\nu, 1\arcsec}$} & \colhead{$1\arcsec$ RMS} & \colhead{3\arcsec} & \colhead{S$_{\nu, 3\arcsec}$} & \colhead{6\arcsec} & \colhead{S$_{\nu, 6\arcsec}$} \\  &  & &  &  & Type & M$_{\odot}$ & & erg cm$^{-2}$ s$^{{-1}}$ & erg cm$^{-2}$ s$^{{-1}}$ & $\arcsec \times \arcsec, \arcdeg$ & Morphology & mJy & $\mu$Jy beam$^{-1}$ & Morphology & mJy & Morphology & mJy }
\colnumbers
\startdata 
6       &       Mrk 335                 & 00 06 19.5    &       +20 12 10.6     &       0.026   &       Sy 1    &       7.23    &       -1.17   &       11.20E-12       &       16.20E-12    & $0.97\arcsec \times 0.87\arcsec$, 71.6\arcdeg &   1       &       1.109   $\pm$   0.007   & 16.6  & 1     &       1.151   $\pm$   0.008   &       1       &       1.233   $\pm$   0.012   \\
13      &       2MASX J00253292+6821442 & 00 25 32.3    &       +68 21 45.3     &       0.012   &       Sy 2    &       7.68    &       -1.92   &       42.60E-12       &       35.40E-12    & $1.57\arcsec \times 0.82\arcsec$, 83.2\arcdeg &       1       &       1.05    $\pm$   0.18    & 17.3  & 1     &       1.696   $\pm$   0.14    &       1       &       1.69    $\pm$   0.16   \\
22      &       CGCG 535-012             & 00 36 20.9   &       +45 39 53.7     &       0.048   &       Sy 1    &       7.36    &       -0.79   &       6.00E-12        &       14.90E-12    & $1.1\arcsec \times 0.83\arcsec$, -74.7\arcdeg &       1       &       0.199   $\pm$   0.021   & 10.7  & 1     &       0.22    $\pm$   0.007   &       1       &       0.26    $\pm$   0.011   \\
28      &       NGC 235A                 & 00 42 52.8   &       -23 32 27.7     &       0.022   &       Sy 1.9  &       8.49    &       -2.06   &       14.70E-12       &       52.20E-12    & $1.85\arcsec \times 0.85\arcsec$, 16.4\arcdeg &   2   &       3.172   $\pm$   0.047   & 54.1  & 1     &       4.259   $\pm$   0.029   &       2       &       4.365   $\pm$   0.01   \\
31      &       MCG-02-02-095            & 00 43 08.8   &       -11 36 03.8     &       0.019   &       Sy 2    &       6.82    &       -1.27   &       2.40E-12        &       9.00E-12     & $1.24\arcsec \times 0.78\arcsec$, 11.4\arcdeg &   0       &       \nodata $\pm$   \nodata & 17.1  & 0     &       \nodata $\pm$   \nodata &       0       &       \nodata $\pm$   \nodata   \\
\enddata
\tablecomments{Physical properties along with the 22 GHz observations of our BAT AGN sample. Columns are (1) \textit{Swift-BAT} 70-month hard X-ray ID, (2) Common identifier, (3) Right Ascension, (4) Declination, (5) Redshift, (6) Seyfert Type, (7) log Black hole mass, (8) log Eddington Ratio, (9) Hard X-ray intrinsic flux (2-10 keV), (10) Ultra-hard X-ray intrinsic flux (14-195 keV), (11) Restoring beam major and minor axes and the position angle, (12) $1\arcsec$ radio morphology class, (13) $1\arcsec$ flux density, (14) $1\arcsec$ rms sensitivity of the image, (15)  $3\arcsec$ radio morphology class, (16) $3\arcsec$ flux density, (17) $6\arcsec$ radio morphology class, and (18) $6\arcsec$ flux density. The morphology classes are represented by the following: 0 $-$ no detection, 1 $-$ compact source, 2 $-$ extended structure, 3 $-$ jet-like or biconical structure (details in Section \ref{sec:morphologies}) .  A machine-readable version of this table is available.}
\end{splitdeluxetable*}

\begin{deluxetable*}{cccccccc}
\tablecaption{22 GHz B-Array Flux and Morphology Measurements \label{tab:22GHz-barray}}
\tablehead{
\colhead{BAT ID} & \colhead{Name} & \colhead{RA} & \colhead{Dec} & B$_{\text{maj}}$$\times$B$_{\text{min}}$, PA & \colhead{0.\arcsec3} & \colhead{S$_{\nu, 0.\arcsec3}$} & \colhead{$0.3\arcsec$ RMS} \\  & &  & & $\arcsec \times \arcsec, \arcdeg$ & Morphology & mJy & $\mu$Jy beam$^{-1}$ }
\colnumbers
\startdata 
28	 &	NGC 235A	  &	00 42 52.8	&	-23 32 27.7	& $0.53\arcsec \times 0.26\arcsec$, 0.35\arcdeg  &	2	&	1.904	$\pm$	0.009 &	26.3 \\
60	 &	Mrk 975	      &	01 13 51.0	&	+13 16 18.2	& $0.34\arcsec \times 0.26\arcsec$, 63\arcdeg    &	1	&	1.167	$\pm$	0.002 &	15.1 \\
116	 &	Mrk 590	      &	02 14 33.6	&	-00 46 00.2	& $0.35\arcsec \times 0.27\arcsec$, 41.1\arcdeg  &	1	&	3.733	$\pm$	0.019 &	35.3 \\
171	 &	MCG+00-09-042 &	03 17 02.2	&	+01 15 18.0	& $0.31\arcsec \times 0.26\arcsec$, 13.7\arcdeg  &	3	&	2.655	$\pm$	0.008 &	23.7 \\
404	 &	Mrk 1210	  &	08 04 05.7	&	+05 06 49.8	& $0.51\arcsec \times 0.26\arcsec$, -54\arcdeg   &	1	&	14.184	$\pm$	0.055 &	100.7\\
470	 &	CGCG 122-055  &	09 42 04.8	&	+23 41 07.0	& $0.66\arcsec \times 0.26\arcsec$, -64.3\arcdeg &	1	&	1.189	$\pm$	0.016 &	14.9 \\
530	 &	NGC 3516	  &	11 06 47.4	&	+72 34 07.3	& $0.33\arcsec \times 0.23\arcsec$, 13.6\arcdeg  &	2	&	1.594	$\pm$	0.016 &	15.8 \\
532	 &	IC 2637	      &	11 13 49.7	&	+09 35 10.6	& $0.42\arcsec \times 0.26\arcsec$, -57.7\arcdeg &	2	&	2.228	$\pm$	0.02  &	24.7 \\
608	 &	Mrk 766	      &	12 18 26.5	&	+29 48 46.7	& $0.43\arcsec \times 0.25\arcsec$, -74.8\arcdeg &	2	&	5.206	$\pm$	0.034 &	36.5 \\
666	 &	UGC 8327	  &	13 15 17.3	&	+44 24 26.0	& $0.31\arcsec \times 0.26\arcsec$, -40.1\arcdeg &	2	&	2.988	$\pm$	0.037 &	22.5 \\
697	 &	Mrk 279	      &	13 53 03.4	&	+69 18 29.4	& $0.33\arcsec \times 0.27\arcsec$, -17.5\arcdeg &	1	&	1.092	$\pm$	0.055 &	15.3 \\
735	 &	Mrk 817	      &	14 36 22.1	&	+58 47 39.4	& $0.49\arcsec \times 0.25\arcsec$, 79.1\arcdeg  &	1	&	2.175	$\pm$	0.006 & 25.7 \\
738	 &	Mrk 477	      &	14 40 38.1	&	+53 30 15.8	& $0.45\arcsec \times 0.25\arcsec$, 82.8\arcdeg  &	2	&	5.186	$\pm$	0.036 & 52.7 \\
942	 &	NGC 6552	  &	18 00 07.3	&	+66 36 54.4	& $0.35\arcsec \times 0.24\arcsec$, 52.8\arcdeg  &	2	&	4.546	$\pm$	0.015 &	20.1 \\
960	 &	UGC 11185	  &	18 16 09.4	&	+42 39 23.0	& $0.34\arcsec \times 0.25\arcsec$, 84.2\arcdeg  &	1	&	0.042	$\pm$	0.002 &	22.3 \\
960	 &	MCG+07-37-031 &	18 16 11.6	&	+42 39 37.3	& $0.34\arcsec \times 0.25\arcsec$, 84.2\arcdeg  &	3	&	1.277	$\pm$	0.003 &	22.3 \\
1183 &	Mrk 926	      &	23 04 43.5	&	-08 41 08.6	& $0.36\arcsec \times 0.24\arcsec$, 10.7\arcdeg  &	3	&	3.366	$\pm$	0.021 &	30.6\\
\enddata
\tablecomments{22 GHz B-Array fluxes and morphology class from \cite{2016ApJ...832..163S}. Columns are (1) \textit{Swift-BAT} 70-month hard X-ray ID, (2) Object name, (3) Right Ascension, (4) Declination, (5)Restoring beam major and minor axes and the position angle, (6)$0.3\arcsec$ morphology class, (7) $0.3\arcsec$ flux density, (8) $0.3\arcsec$ rms sensitivity of the image. The morphology classes are represented by the following: 0 $-$ no detection, 1 $-$ compact source, 2 $-$ extended structure, 3 $-$ jet-like or biconical structure (details in Section \ref{sec:morphologies}).}
\end{deluxetable*}

\subsection{Radio Images: Reduction and Processing} \label{subsec:radio}

The typical time on source was $3-10$ minutes and, along with the 8 GHz bandwidth for the K-band (22 GHz) receiver, allowed us to reach $1\,\sigma$ sensitivities of $\sim\,16\,\mu$Jy. For objects with longer integrations, the on source integration time ranged from 13 to 20 minutes. The 17 objects that were observed in B-array at 22 GHz with $\sim\,0.3$\arcsec~resolution had a 1\,$\sigma$ sensitivity of $\sim\,10\,\mu$Jy. In addition to adding more sources in the second phase of the survey, longer observations were also obtained with the VLA in C-array for objects that were showing tentative extended star-forming emission. The typical $1\,\sigma$ sensitivities for these observations are around $\sim\,8\,\mu$Jy. These observations are listed as additional entries in Table \ref{tab:22GHz_props}. Overall, the majority of our sample was taken with the VLA in C-array, and thus the sample has a sensitivity of $\sim16$ $\mu$Jy with $\sim1\arcsec$ resolution.

We followed the same observing strategy as in Papers I and II: 2-3 science targets were put into 1-hour observing blocks. Each observing block started with X- and K-band attenuation scans. The flux and bandpass calibrations for all three phases used 3C 48, 3C 138, 3C 286, or 3C 147, depending on sky position and antenna wraps. The project codes for the entire survey are listed here: VLA/13A-281, VLA/14B-289, VLA/15A-431, VLA/17A-267, VLA/18B-245, and VLA/20A-158.

The raw data was reduced and processed using the standard VLA reduction pipeline through Common Astronomy Software Applications (CASA) version 6.5.5 \citep{2007ASPC..376..127M, TheCASAteam_2022}\footnote{\url{https://casa.nrao.edu/}} for the third phase of our radio survey. Similarly, we follow the reduction procedure described in Papers I and II. For each observation, all 64 channels were averaged within each spectral window. Since all of our sources were in the phase center of our observations, we expect the bandwidth smearing effects to be minimal near the observed sources. Each object was interactively cleaned using the \texttt{tclean} task in CASA to a threshold of 0.05 mJy (that is, 3$\sigma$ above our noise) with Briggs weighting (\texttt{robust} = 0.5), and the primary beam was corrected by setting \texttt{pblimit} to $-0.001$. After creating an image, we visually inspected for RFI effects; where we considered RFI to be present, we attempted to minimize its impact by flagging and removing the worst-affected spectral windows. We did not significantly lose sensitivity due to RFI flagging, and, often times, we retain most of the 8 GHz bandwidth after flagging the affected spectral windows. In fact, we did not lose any objects due to RFI in the final phase and we recovered a few objects from the previous phases that had been discarded due to RFI. If the source's flux density was greater than 1 mJy or a S/N threshold greater than 50, the objects were phase self-calibrated before final cleaning to further reduce systematics. Altogether, we have detected 227 out of 231 \textit{Swift}-BAT AGN with the VLA in C-array, thus giving a 98\% detection fraction at 22 GHz.

\subsection{Flux Measurements in the Radio} \label{subsec:flux}

We follow the flux measurement procedure mentioned in Papers I and II. To gather the flux densities and flux density errors of the cores of all of our sources, we used the CASA command \texttt{imfit} to fit each Stokes I image component with an elliptical Gaussian. All of the compact sources were well fit by this method. If \texttt{imfit} failed with its fitting procedure, then we would draw an ellipse with the restoring beam's major and minor axes and fit the ellipse with a Gaussian through CASA. This method matches the flux density that the \texttt{imfit} method would return. Often, the images contain extended low-surface-brightness emission. To image this faint emission, we re-cleaned and created 3\arcsec~and 6\arcsec~\text{uv}-tapered Briggs weighted images for every object in our sample. We chose a 6\arcsec~\text{uv}-tapered image to match the \textit{Herschel} beam, which covers a spatial range of $0.21-5.874$ kiloparsecs for our sample. For most of our objects, the 6\arcsec~\textit{uv}-tapered image is unresolved (i.e. there is no extended structure beyond 6\arcsec). Note that the Largest Angular Scale (LAS) for the C-configuration of the VLA is $\sim33\arcsec$ for snapshots, more than 5 times the size of our largest taper.

For Fairall 272 and II Zw 83, we were unable to fit the core in the images, finding only jet-like or binconical morphologies dominating these objects. We do note that one of the lobes in both Fairall 272 and II Zw 83 are centered on the nucleus of their host galaxy. For the purposes of this paper, we only consider and work with the flux densities derived from those nuclei-based lobes. 

For the redshift range of our sample, a 1\arcsec~beam corresponds to a scale of 35 - 987 parsecs. Paper I has shown that while this resolution is high enough to separate out the majority of star formation and outflow-related emission from the AGN core, there is a good chance that there are still unresolved extended radio structures within the beam, especially for the most distant objects in the sample. In fact, \cite{2022ApJ...940...52S} have found that compact star formation occurs on parsec scales. Interestingly, when comparing the flux densities of the $0.3\arcsec, 1\arcsec, 3\arcsec, \text{and}\, 6\arcsec$, we find that flux is resolved out on the smaller scales and is very rarely constant across all four resolutions. This is supported by the fact that we were able to pick up diffuse extended radio structures with the 3\arcsec and 6\arcsec resolution images.

\begin{figure*}
    \begin{center}
    \includegraphics[trim=0mm 0mm 0mm 0mm, clip, width=2.1\columnwidth]{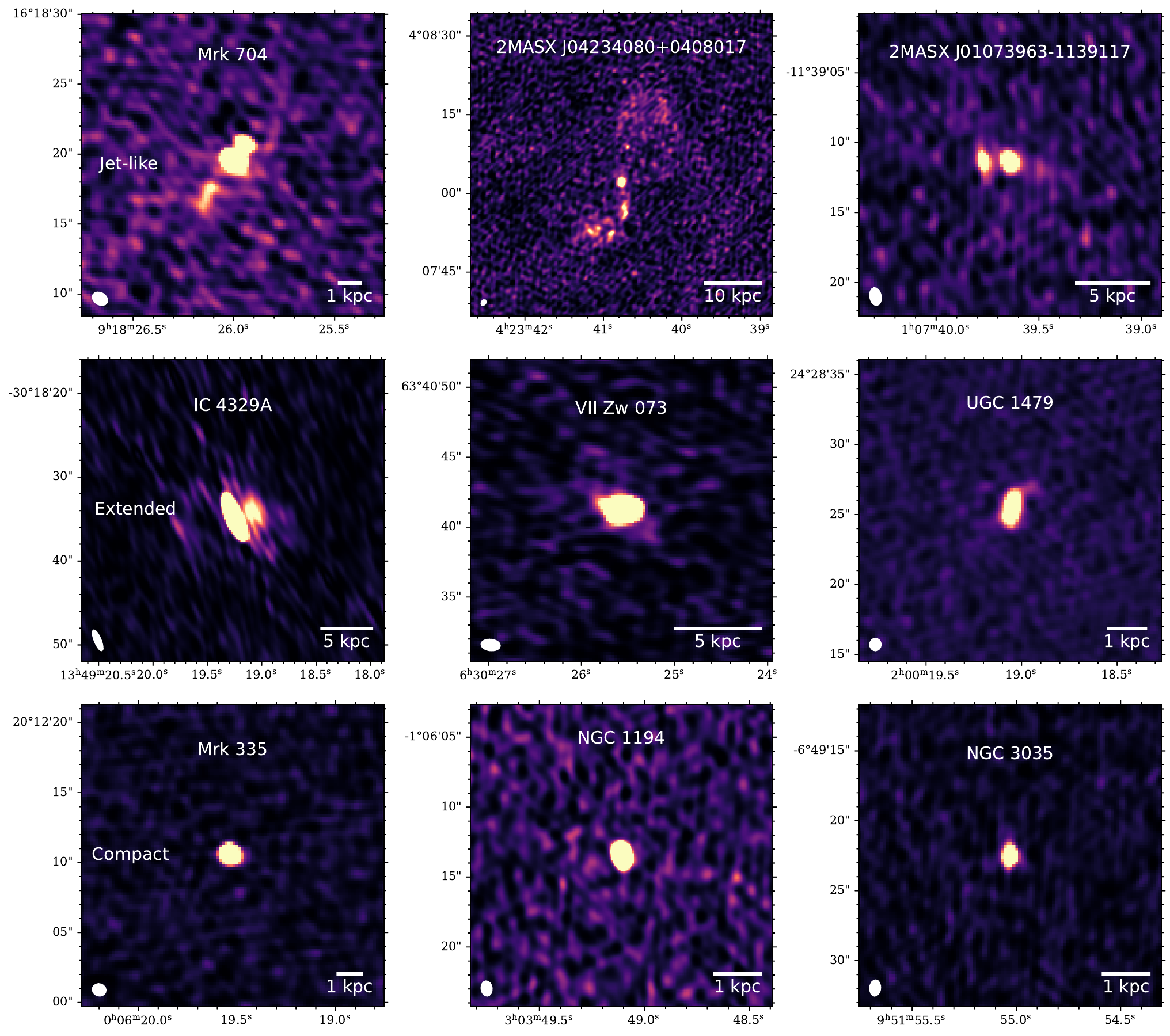}
    \caption{22 GHz images of the BAT AGN with various morphological classes. The top row represents different types of jet-like morphologies (type 3 morphological class). The middle row represents the various extended structures (type 2 morphological class). The bottom row represents the compact core-dominated sources (type 1 morphological class). Each figure includes the beam (white, bottom left), and a scale bar representing either 1, 5, or 10 kiloparsecs.}
    \label{fig:22GHz-morphologies}
    \end{center}
\end{figure*}

\begin{figure*}
    \begin{center}
    \includegraphics[trim=0mm 0mm 0mm 0mm, clip, width=2.1\columnwidth]{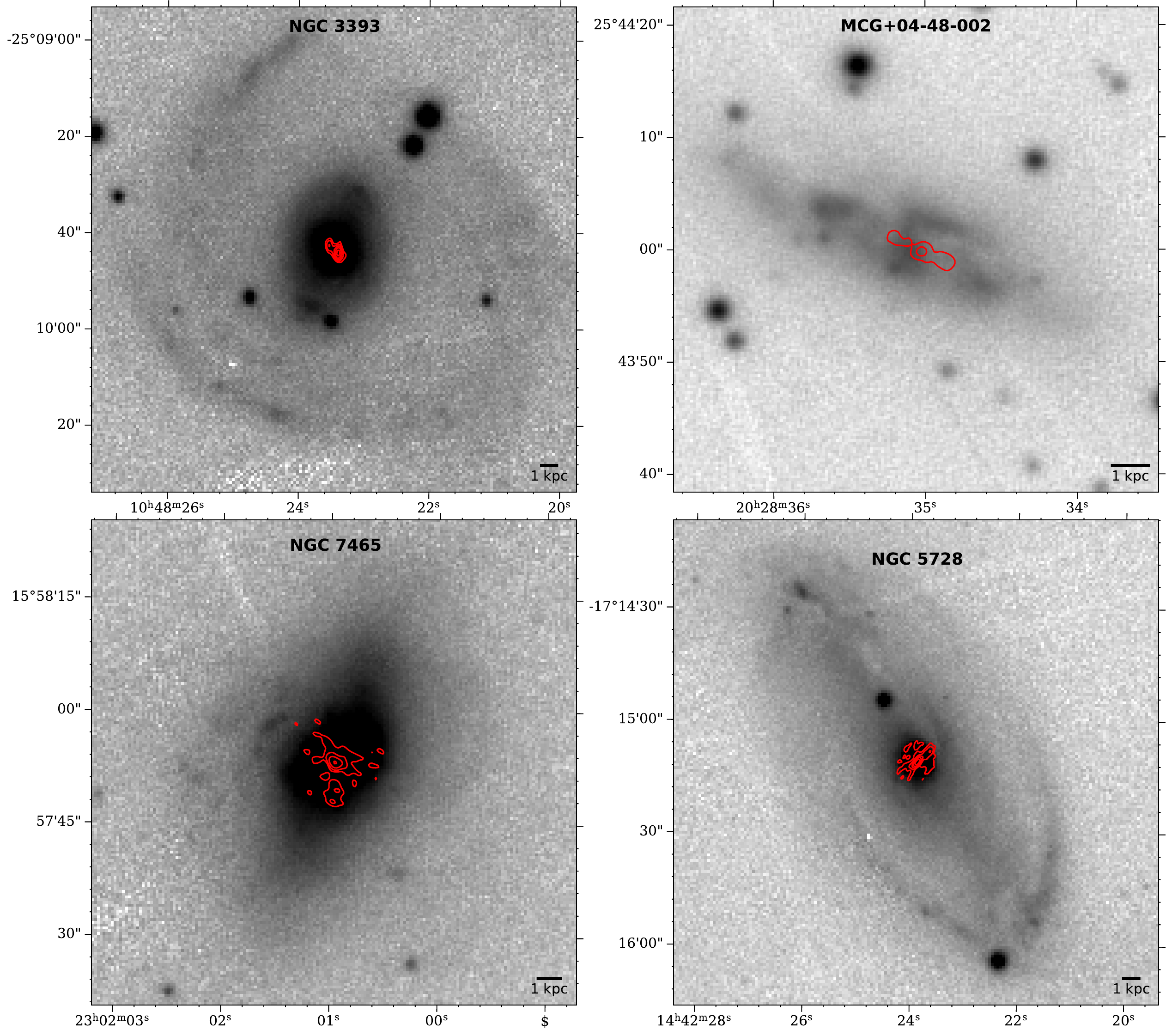}
    \caption{PanSTARRS g-band images with red 1\arcsec~radio contours overlaid on top.}
    \label{fig:22GHz-contours}
    \end{center}
\end{figure*}

\subsection{\texorpdfstring{H$_2$O}.  Maser Emission} \label{subsec:Maser}

Because H$_2$O maser emission occurs near 22 GHz, we need to assess the possible impact of maser emission on our detection fraction and flux densities. By comparing our sample with the Public Water Maser List\footnote{\url{https://safe.nrao.edu/wiki/bin/view/Main/PublicWaterMaserList}}, which is provided through the Megamaser Cosmology Project \citep[e.g.,][]{2017ApJ...834...52G}, we found that 17 sources have known water masers: Mrk\,348, NGC\,1052, NGC\,1106, NGC\,1194, CGCG\,468-002, Mrk\,3, Mrk\,1210, NGC\,3081, NGC\,3079, NGC\,3393, NGC\,4388, NGC\,5506, NGC\,5728, NGC\,6240, NGC\,7479, NGC\,235A, and VII\,Zw\,073. We compared the flux density from the water megamaser emission to our 22 GHz continuum emission. We found that NGC\,3079 has the highest ``contamination'' of water megamaser emission at $1.3\%$ of the 22 GHz continuum flux. Therefore, our data are unlikely to be significantly affected by H$_2$O maser emission.

\section{Radio Morphologies} \label{sec:morphologies}
Our sample has a wide range of observed morphologies at 22 GHz, which we classify through visual inspection into four categories, $0-3$, where $0$ denotes a non-detection. For the non-detections, we only have 4 sources out of total sample (231 sources) that are not detected within our survey. A type $1$ morphological class represents compact sources,  $44.16\%$ (102 sources) of our full sample. For our purposes, a compact source does not show extended emission. A type $2$ morphological class represents sources with extended structures that are not jet-like or biconical; these sources make up $45.89\%$ (106 sources) of our sample. These extended structures are likely associated with star formation, but may also be caused by asymmetric shocks from an AGN-driven outflow \citep[e.g.,][]{2022AJ....164..122M}. Finally, a type $3$ morphological class represents jet-like or biconical sources. These sources make up $8.23\%$ (19 sources) of our sample. We have defined the jet-like morphology to be linear and symmetric about an axis. These jet-like radio morphologies could also very well be jetted outflows or they could be an AGN wind or possibly dual AGN. Figure \ref{fig:22GHz-morphologies} shows an example for each morphological class except for the non-detections. In Figure \ref{fig:22GHz-contours}, our 1\arcsec~extended radio contours are overlaid on PanSTARRS g-band images \citep{2020ApJS..251....7F} of four BAT AGN, highlighting the fact that we are probing the inner nuclei of these galaxies.

Linear kiloparsec-scale morphologies can indicate a young or frustrated jet, but this may not be their most likely origin. \cite{2023ApJ...953...87F} reported that such linear jet-like morphologies could instead indicate an AGN wind that causes a symmetrical shock, which resembles a jet. Because of this, unlike Papers I and II, we do not use lower-frequency surveys to help classify objects as jet-like, and we group by 22 GHz morphology only, not speculating on jetted vs. wind activity. 

The origin of the radio emission present in all of our morphologies will be discussed in more detail in Section \ref{subsec:Discussion}. We will disentangle the AGN emission from unresolved star formation using detailed radio SED modeling in a forthcoming paper.

\section{Results} \label{sec:Results}
\subsection{The Coronal Radio-X-ray Correlation} \label{subsec:Coronal}

\begin{figure}[htpb!]
\begin{center}
    \includegraphics[trim=0mm 0mm 0mm 0mm, clip, width=\columnwidth]{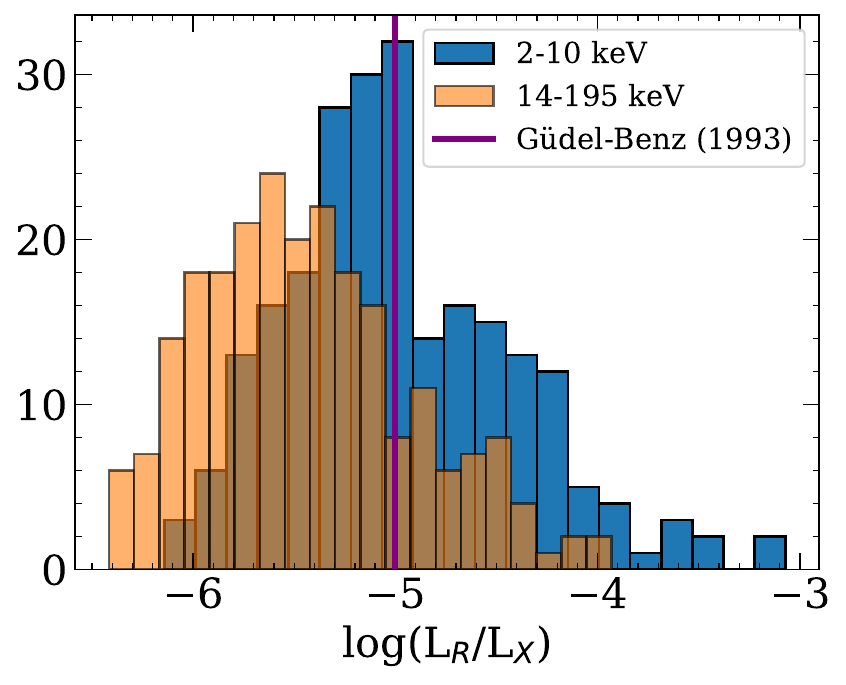}
    \caption{Histogram of the ratio of the 22 GHz radio luminosity to the X-ray luminosity. The magenta line represents \cite{1993ApJ...405L..63G}'s derived value for coronally active stars ($10^{-5}$).}
    \label{fig:22GHz-coronal-ratio}
    \end{center}
\end{figure}

In this section, we compare our sample to  the $L_R$/$L_X$ relation first established by \cite{1993ApJ...405L..63G} for stellar coronae: $L_R/L_X \sim 10^{-5}$. Later, \cite{2008MNRAS.390..847L} found that this relation also held for radio-quiet quasars, suggesting that the origin of radio emission in the GHz regime for radio-quiet AGN is not an unresolved jet, but is instead a population of electrons related to the coronal X-ray emission. Both Papers I and II found that the sample followed the \cite{1993ApJ...405L..63G} $L_R/L_X$ relation for coronally active stars. Figure \ref{fig:22GHz-coronal-ratio} shows the ratio of radio to X-ray luminosities for the full sample presented in this paper. A magenta line has been plotted showing \cite{1993ApJ...405L..63G}'s derived $L_R/L_X$ value for coronally active stars ($10^{-5}$). Furthermore, the $L_R/L_X$ median and mean value for the hard X-rays (2-10 keV) are centered around $\sim-5$, while the $L_R/L_X$ for ultra-hard (14-195 keV) X-rays are centered around $\sim-5.5$. Both of these values are consistent with the \cite{1993ApJ...405L..63G} $L_R/L_X$ relation for coronally active stars.

Following the procedure of Paper II, we control for the mutual dependence on redshift by calculating the partial Kendall's-$\tau$ correlation coefficient using the method of \cite{1996MNRAS.278..919A} for the radio and X-ray luminosities.  Our results are shown in Table \ref{tab:22GHz-kendall-tau}. For the partial Kendall's-$\tau$ analysis, we have elected to also include multiple radio observations of objects (see Section \ref{subsec:repeat}), since we do not have simultaneous radio and X-ray data. Furthermore, the variability in the radio can indicate different states of a given AGN. This results in a sample of 231 observations. After accounting for the redshift, our sample shows a significant correlation between radio and both hard and ultra-hard X-ray in both flux and luminosity. For both the hard and ultra-hard X-ray luminosities, we found a Kendall's $\tau$ statistic $\sim0.33$ where both had a p-value $<< 0.01$. With this p-value, the null hypothesis (P$_{\text{null}}$), which is the probability that a correlation is not present, is extremely small. Therefore, our results state that a correlation is almost certainly present even once the influence of redshift is accounted for. This is consistent with previous results \citep[e.g.,][]{2022A&A...668A.133D}.

Since we do not have to account for the redshift for our radio and X-ray fluxes, we instead used the Python package \verb|scipy.stats.kendalltau| to find the Kendall's $\tau$ statistic and corresponding p-value. It is important to analyze the fluxes because the strongly correlated luminosity-luminosity relations could easily have been induced artificially through the redshift \citep[e.g.,][]{2019ApJ...877...63S}. By analyzing the fluxes, we can see if the luminosity-luminosity correlations hold or if they fall apart in flux-flux space. We found a Kendall's $\tau$ statistic of $0.331$ with a p-value $<< 0.01$ for the ultra-hard X-rays, and a Kendall's $\tau$ statistic of $0.287$ with a p-value of $<< 0.01$ for the hard X-rays. Once again, the probability of the null hypothesis is extremely small, which implies that a correlation is almost certainly present. Our results here are also listed in Table \ref{tab:22GHz-kendall-tau}.

\begin{table}
\centering 
\begin{tabular}{ c c c c c c c c }
 \hline
 \multicolumn{7}{c}{Full Sample} \\
 \hline
 X & Y & Z & N & $\tau$ & $\sigma$ & P$_{\text{null}}$\\
 \hline
 $L_{HX}$  & $L_R$ & $z$ & 231 & 0.336 & 0.031 & $<<$ 0.01\\
 
 $L_{UHX}$ & $L_R$ & $z$ & 231 & 0.330 & 0.031 & $<<$ 0.01 \\
 \hline \hline 
 X & Y & Z & N & $\tau$ & $\sigma$ & P$_{\text{null}}$\\
 \hline 
 $F_{HX}$  & $F_R$ & $$\nodata$$ & 231 & 0.287 & $$\nodata$$ & $<<$ 0.01\\
 $F_{UHX}$ & $F_R$ & $$\nodata$$ & 231 & 0.331 & $$\nodata$$ & $<<$ 0.01 \\
 \hline
\end{tabular}
\caption{Top: Results of the partial correlation analysis for the hard X-ray luminosity ($L_{HX}$), the ultra-hard X-ray luminosity ($L_{UHX}$), radio luminosity ($L_{R}$), and redshift ($z$). Bottom: Results of the Kendall's $\tau$ correlation analysis for the hard X-ray flux ($F_{HX}$), the ultra-hard X-ray flux ($F_{UHX}$), and radio flux ($F_{R}$). The columns are: (1) X is the independent variable, (2) Y is the dependent variable, (3) Z is the influencing variable, (4) N is the total number of objects in our sample, (5) Top: The partial Kendall's $\tau$ coefficient calculated using the method of \cite{1996MNRAS.278..919A} Bottom: The Kendall's $\tau$ statistic calculated using scipy.stats.kendalltau, (6) $\sigma$, the square root of the variance of $\tau$, and (7) Top: the probability, P$_{\text{null}}$, of accepting the null hypothesis that X and Y are uncorrelated once the influence of Z is accounted for. Bottom: the probability, P$_{\text{null}}$, of accepting the null hypothesis that X and Y are uncorrelated.}
\label{tab:22GHz-kendall-tau}
\end{table}

\begin{figure*}[htbp!]
    \centering 
    (a) Optical Morphology\par\medskip
    \includegraphics[trim=0mm 0mm 0mm 0mm, clip, width=2\columnwidth]{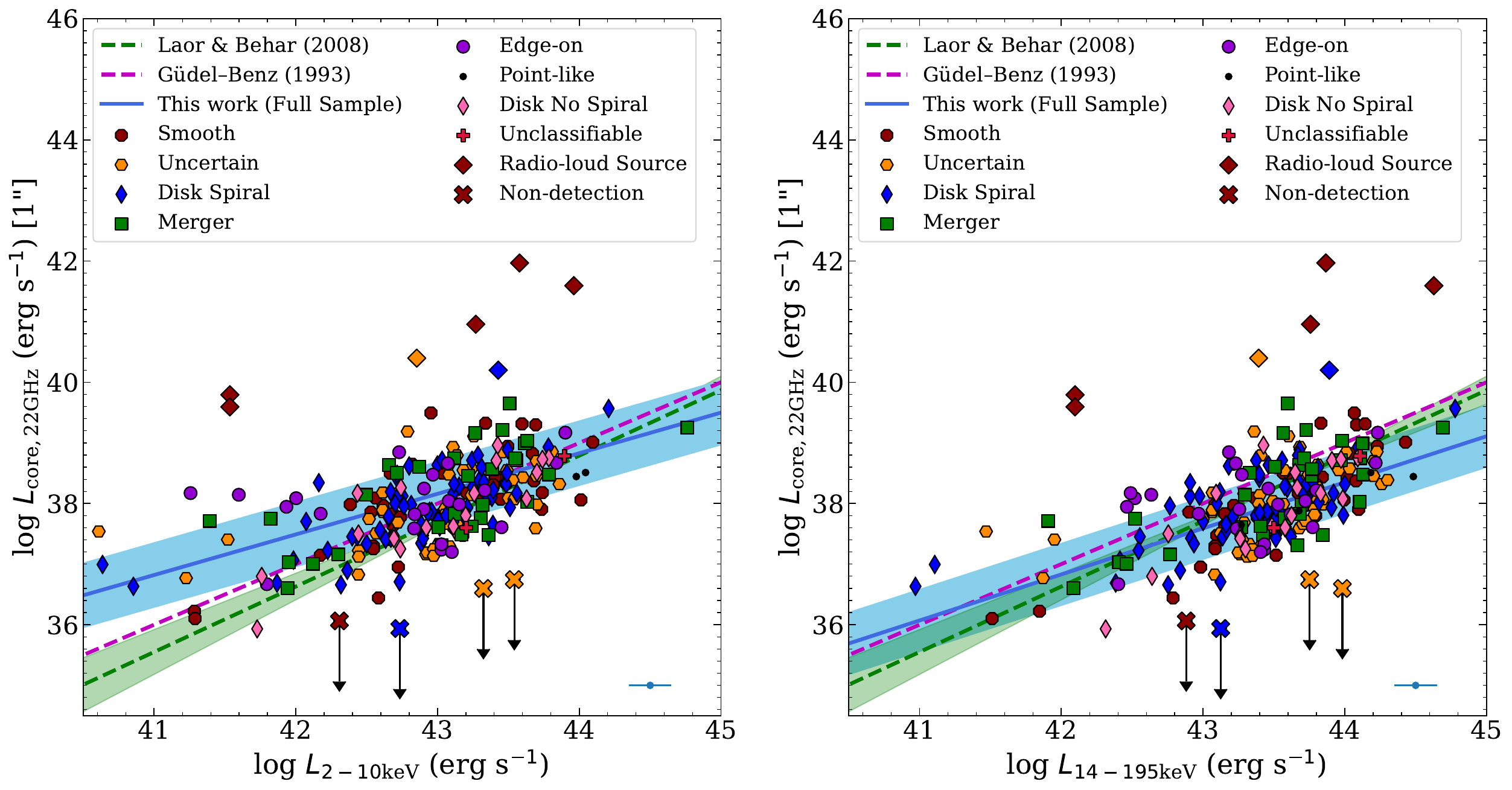} \\
    \quad
    \centering
    \centering
    (b) Spectral Type\par\medskip
    \includegraphics[trim=0mm 0mm 0mm 0mm, clip, width=2\columnwidth]{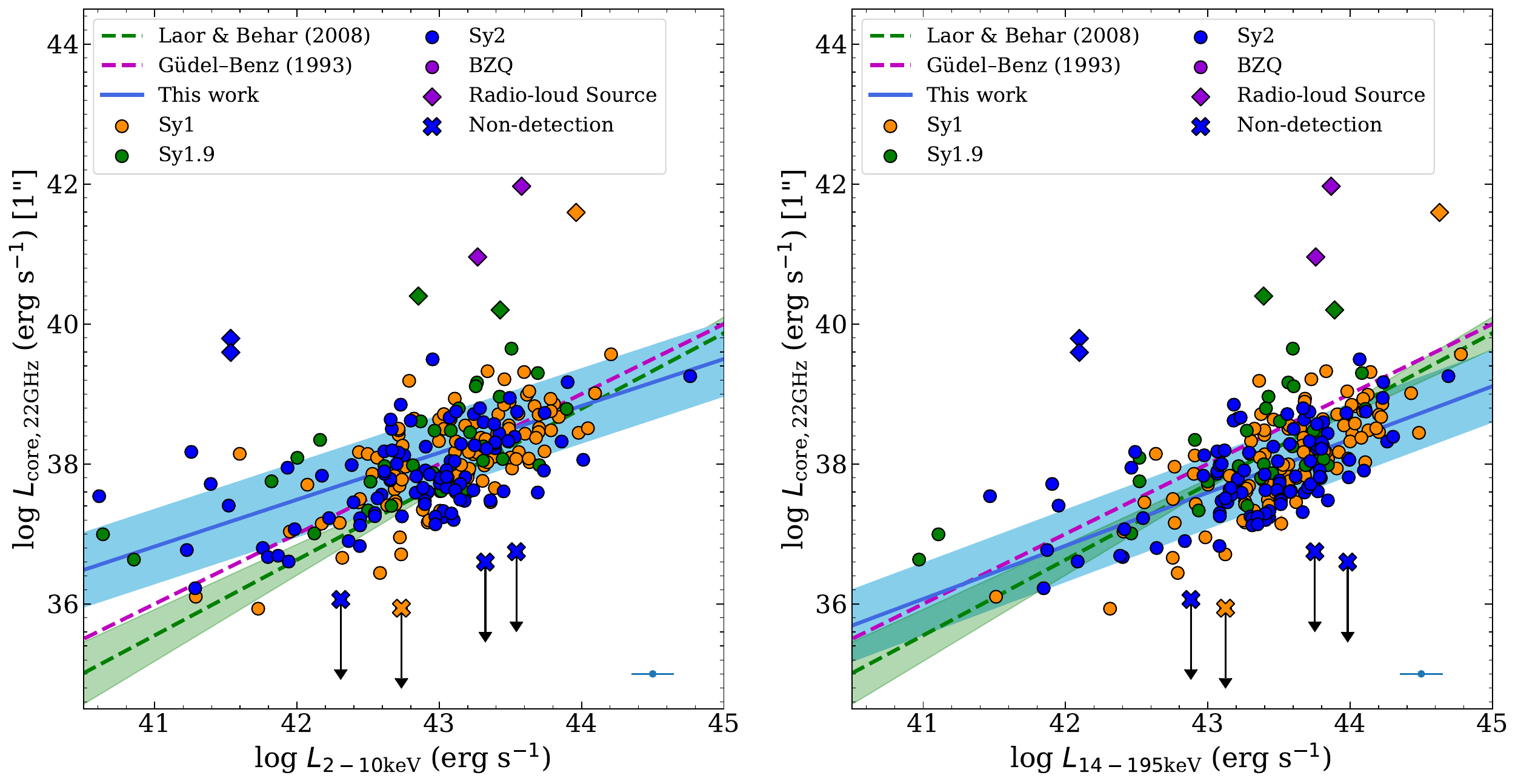} \\
    \caption{22 GHz radio luminosity versus hard X-ray (2-10 keV) luminosity (left) and ultra-hard X-ray (14-195 keV) luminosity (right). Typical error bars are shown on the bottom right. The black arrows represent the upper limits of three non-detections in our sample with 70-month \textit{Swift}-BAT data. The dark blue line shows the linear regression calculated using dynesty. The blue shaded region represents the scatter of the data. The purple dashed line shows the $L_R$/$L_X$ relation for stellar coronae from \cite{1993ApJ...405L..63G}. The green dashed line (and associated error region) represents the minimized scatter in the radio for the $L_R$/$L_X$ from \cite{2008MNRAS.390..847L}. Radio-loud sources have been plotted with diamonds.}
    \label{fig:22GHz-coronal-relation}
\end{figure*}

\begin{figure*}[htbp!]
    \centering
    (a) 22 GHz radio morphology\par\medskip
    \includegraphics[trim=0mm 0mm 0mm 0mm, clip, width=2.1\columnwidth]{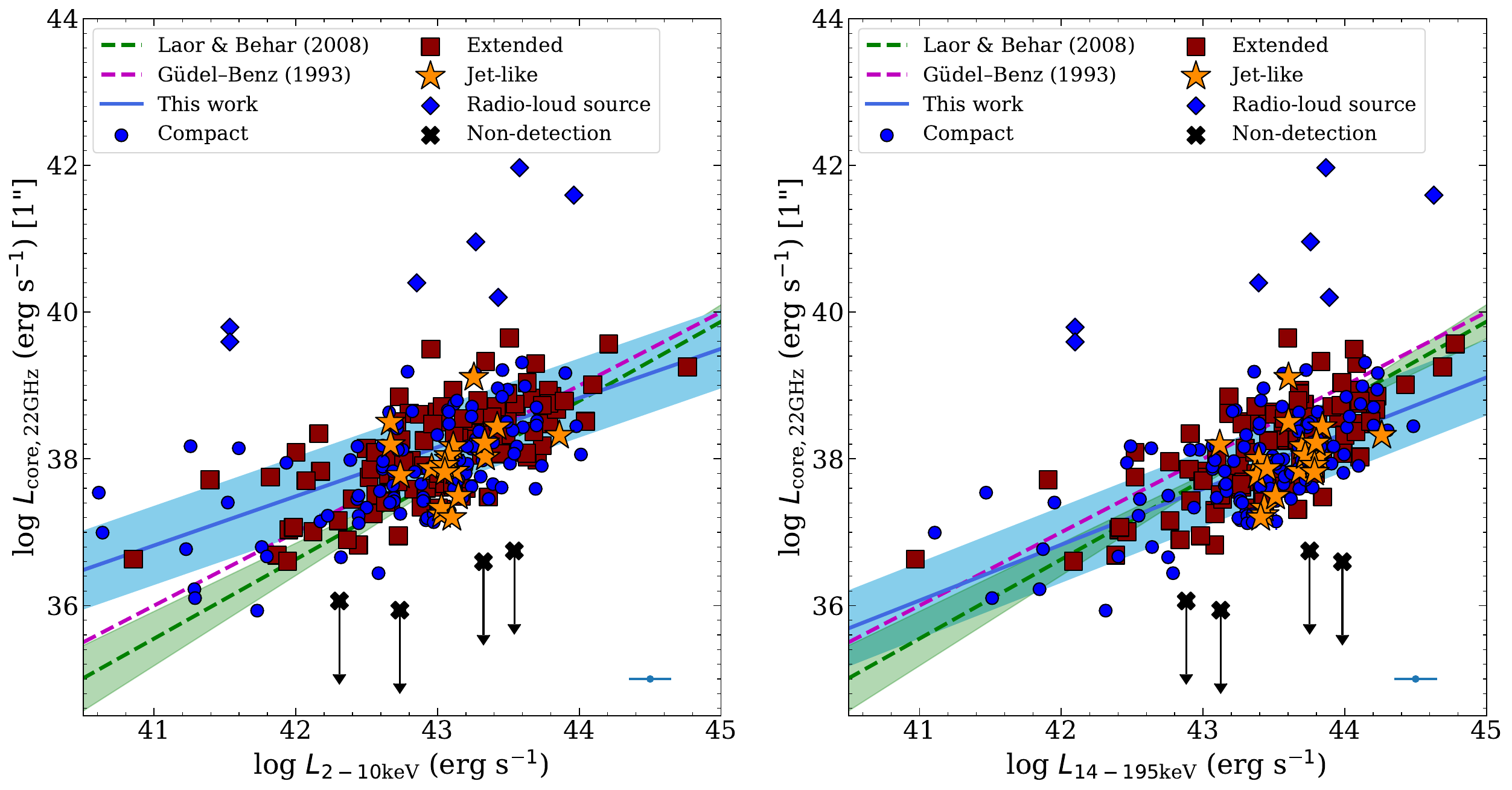} \\
    \centering
    \quad
    (b) Eddington ratio\par\medskip
    \includegraphics[trim=0mm 0mm 0mm 0mm, clip, width=2.2\columnwidth]{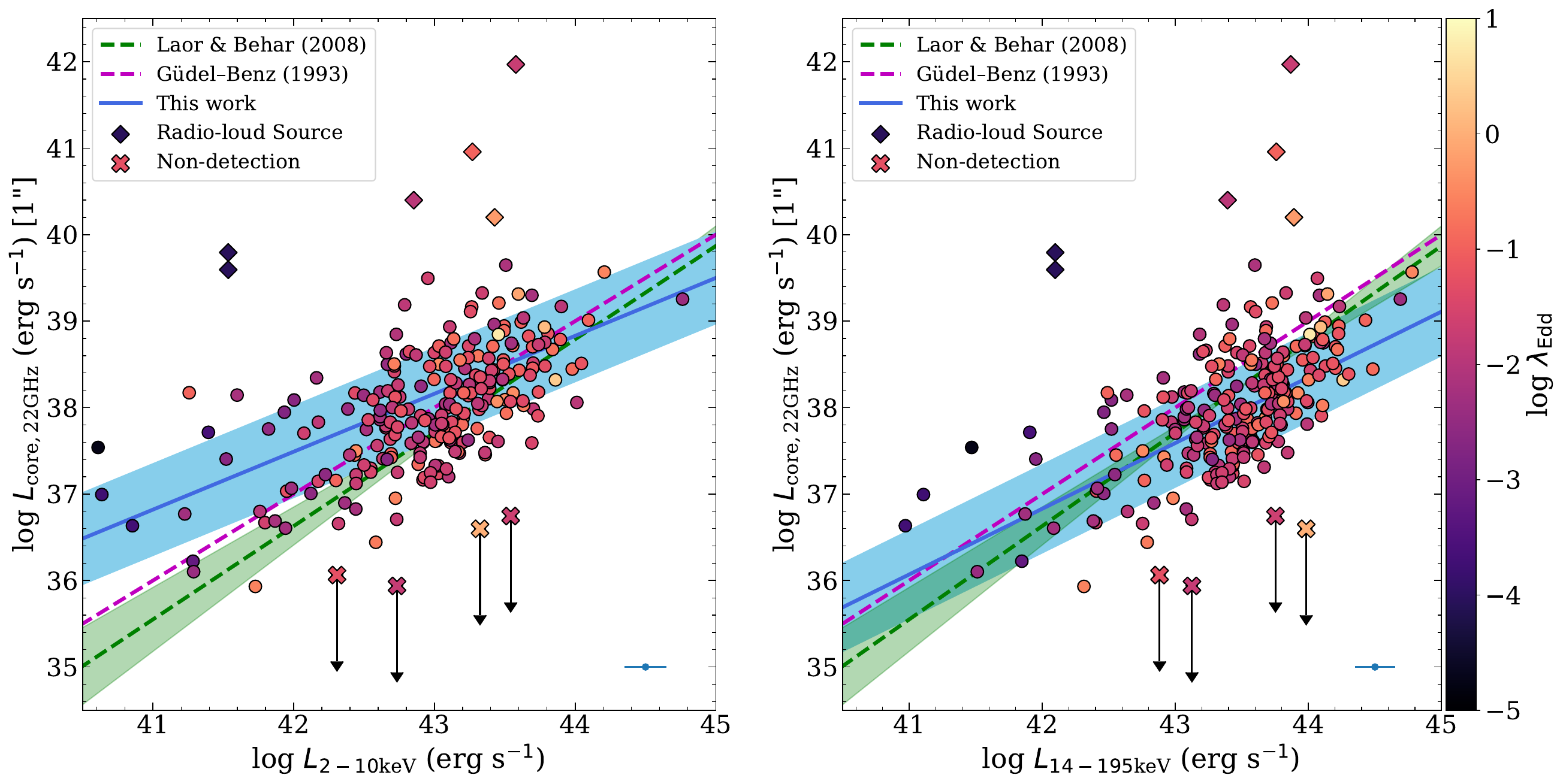} \\
    \centering
  \caption{22 GHz radio luminosity versus hard X-ray (2-10 keV) luminosity (left) and ultra-hard X-ray (14-195 keV) luminosity (right). Typical error bars are shown on the bottom right. The black arrows represent the upper limits of three non-detections in our sample with 70-month \textit{Swift}-BAT data. The dark blue line shows the linear regression calculated using dynesty. The blue shaded region represents the scatter of the data. The purple dashed line shows the $L_R$/$L_X$ relation for stellar coronae from \cite{1993ApJ...405L..63G}. The green dashed line (and associated error region) represents the minimized scatter in the radio for the $L_R$/$L_X$ from \cite{2008MNRAS.390..847L}. Radio-loud sources have been plotted with diamonds.}
  \label{fig:22GHz-coronal-relation2}
\end{figure*}

Figures \ref{fig:22GHz-coronal-relation} and \ref{fig:22GHz-coronal-relation2} compare the 22 GHz radio luminosities of the full sample to both the hard (2-10 keV) and ultra-hard (14-195 keV) X-rays including non-detections as radio upper limits. We did not find any distinct populations in our plot when we included the galaxy's spectral type, optical morphology, or radio morphology at 22 GHz, but we do see a trend with the Eddington ratio. The galaxy's spectral type and Eddington ratio are from \cite{Koss_2022}, and the optical morphologies are from Parra Tello et al. (2025, in preparation). We used the nested sampling package \verb|dynesty| in Python \citep{2020MNRAS.493.3132S, sergey_koposov} for a linear regression of the form 
\begin{equation}
Y = mX + b + \epsilon
\end{equation}
where $m$ is the slope, $b$ is the intercept, and $\epsilon$ is the intrinsic scatter of the sample. \verb|Dynesty| is based on the nested sampling techniques of \cite{2004AIPC..735..395S, 10.1214/06-BA127} along with the dynamic nested sampling of \cite{higson_2019}. These techniques are optimized for the inference of posterior distributions using the bounding method of \cite{2009MNRAS.398.1601F}.

We do not include here the coronally active stars and stellar mass black holes that are present in both \cite{1993ApJ...405L..63G} and \cite{2008MNRAS.390..847L} because of the difference in observed frequencies (5 GHz vs 22 GHz). As for extrapolating to lower frequencies from 22 GHz, we decided that this may not be the best course of action given that we see 22 GHz excess that deviates from a synchrotron power-law \citep{2018MNRAS.478..399B}. See Section~\ref{subsec:Discussion} for more information. 


From the linear regression, we found the following relation for the 22 GHz and ultra-hard X-rays (14-195 keV) luminosities: 
\begin{equation}
\text{log L}_{\text{22 GHz}} = 4.87^{+4.99}_{-4.93} + 0.76^{+0.11}_{-0.11} \ \text{log L}_{\text{14-195 keV}}
\end{equation}
with an intrinsic scatter of $0.51^{+0.05}_{-0.04}$ dex. For the 22 GHz and hard X-rays (2-10 keV) luminosities, we found:
\begin{equation}
\text{log L}_{\text{22 GHz}} = 9.41^{+4.27}_{-4.71} + 0.67^{+0.11}_{-0.10} \ \text{log L}_{\text{2-10 keV}}
\end{equation}
with an intrinsic scatter of $0.53^{+0.05}_{-0.04}$ dex. 

The offset in the luminosity ratio shown in Figure \ref{fig:22GHz-coronal-ratio} manifests itself in the differing slopes and normalizations of the luminosity relations shown in Figures \ref{fig:22GHz-coronal-relation} and \ref{fig:22GHz-coronal-relation2}.

\begin{figure*}[htbp!]
    \begin{center}
    \includegraphics[trim=0mm 0mm 0mm 0mm, clip, width=2.2\columnwidth]{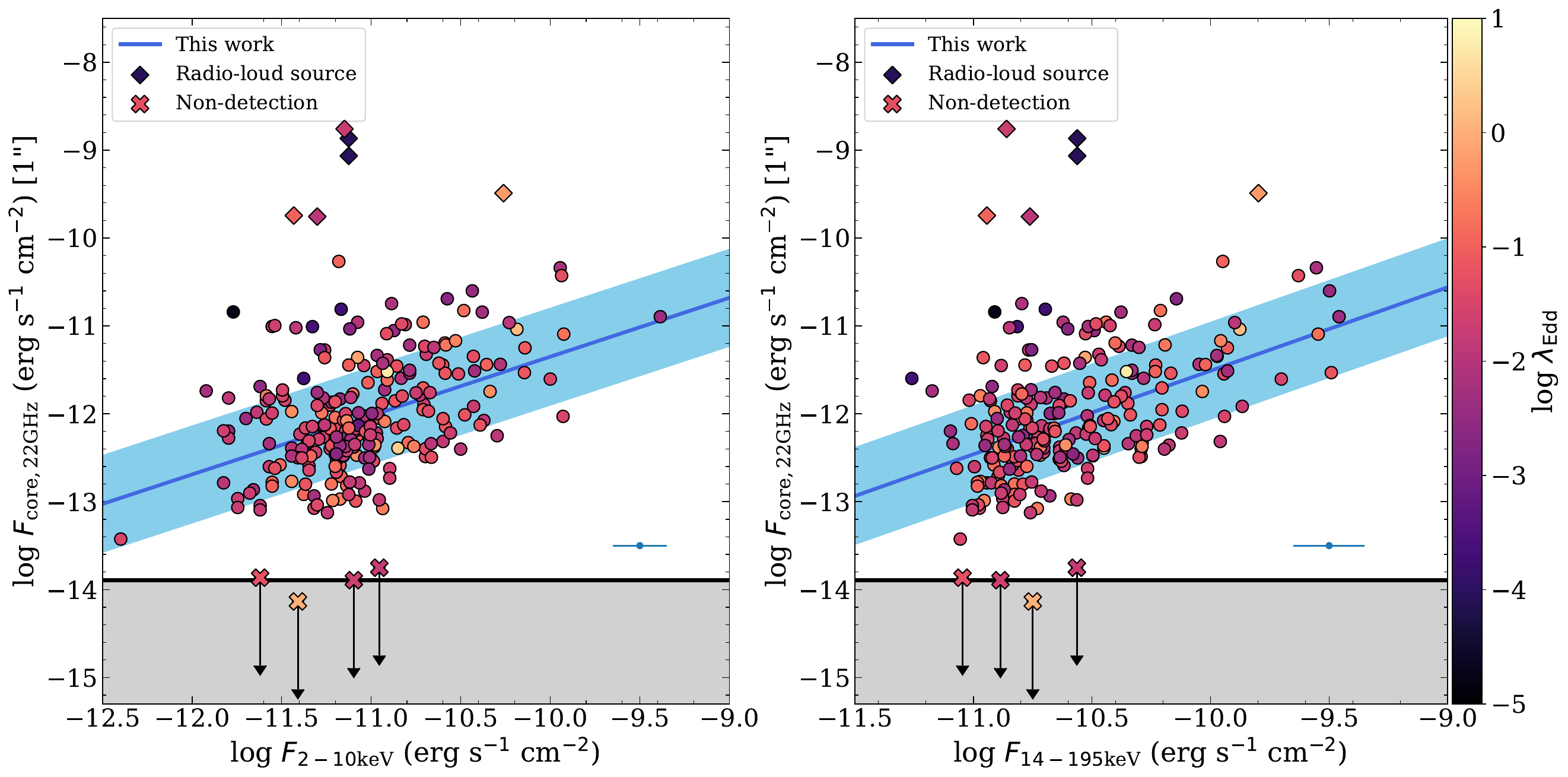}
    \caption{22 GHz radio flux versus hard X-ray (2-10 keV) flux (left) and ultra-hard X-ray (14-195 keV) flux (right). The color bar and the colors represent the Eddington Ratio for each source. Typical horizontal error bars are shown in the bottom right; radio flux errors are smaller than the points. The black arrows represent the upper limits of three non-detections in our sample. The dark blue line shows the linear regression calculated using dynesty. The blue shaded region represents the scatter of the data. The black line and associated shaded region represents the sensitivity limit ($\sim16\,\mu\text{Jy}$) for our survey. Radio-loud sources have been plotted with diamonds. }
    \label{fig:22GHz-coronal-flux}
    \end{center}
\end{figure*}

Figure \ref{fig:22GHz-coronal-flux} illustrates a relation between our 22 GHz fluxes and our hard and ultra-hard X-ray fluxes. From the linear regression, we found the following relation for the 22 GHz and ultra-hard X-rays (14-195 keV) fluxes: 
\begin{equation}
\text{log F}_{\text{22 GHz}} = -2.00^{+2.12}_{-2.11} + 0.95^{+0.20}_{-0.20} \ \text{log F}_{\text{14-195 keV}}
\end{equation}
with an intrinsic scatter of $0.53^{+0.05}_{-0.04}$ dex. For the 22 GHz and hard X-rays (2-10 keV) fluxes, we found:
\begin{equation}
\text{log F}_{\text{22 GHz}} = -4.65^{+1.85}_{-1.86} + 0.67^{+0.17}_{-0.17} \ \text{log F}_{\text{2-10 keV}}
\end{equation}
with an intrinsic scatter of $0.55^{+0.05}_{-0.05}$ dex. For both the luminosity and flux linear regressions, the X-ray fluxes and errors for our sources are described in Paper II.
 
Our complete sample is broadly consistent with the \cite{1993ApJ...405L..63G} and \cite{2008MNRAS.390..847L} relations. The vast majority of the sample has Eddington ratios less than $10^{-2}$. Limiting the sample to sources with Eddington ratios less than $10^{-3}$ has no effect on our linear regressions results. The sources that fall well above the $L_R$/$L_X$ relation are all known radio-loud sources that we mentioned in Section \ref{sec:sample} (diamonds in Figures \ref{fig:22GHz-coronal-relation} and \ref{fig:22GHz-coronal-relation2}): Arp\,102B, Mrk\,348, NGC\,1052 (2 observations), PKS\,2331-240, and 2MASX\,J23272195+1524375. All of these sources had integrated flux densities at 22 GHz greater than 200 mJy, lumninosities greater than $10^{39}$ erg $\text{s}^{-1}$, and have clear jet-like morphologies in other surveys \citep[e.g.][]{2019MNRAS.488.4317B, 2020ApJ...904...83S, 2024arXiv240206166K}; therefore, it follows that they will lie systematically above the $L_R$/$L_X$ relation due to excess synchrotron emission originating from the jet. Additionally, three out of the four non-detections in our sample had X-ray data from the 70-month \textit{Swift}-BAT survey. Plotting these sources in Figures \ref{fig:22GHz-coronal-relation}, \ref{fig:22GHz-coronal-relation2}, and \ref{fig:22GHz-coronal-flux} shows that these non-detections may indicate a possible `radio-silent' population, as they are not anomalously faint in the X-rays. 

\cite{2020MNRAS.495.3943C} and \cite{2022MNRAS.515..473P} have constructed a similar X-ray-detected sample of 88 and 89 AGN, respectively, with VLA radio follow-up from 5-45 GHz. They found that their sources, excluding radio loud sources, follow a similar $L_R$/$L_X$ relation across all frequencies. VLBA analogs to our survey at 1, 4, 6, 8, and 22 GHz can be found in \cite{2021ApJ...906...88F} and \cite{2022ApJ...936...76S}. \cite{2021ApJ...906...88F} found that their sample primarily falls below the \cite{1993ApJ...405L..63G} relation, and their non-detections point to a possible population of AGN that do not have any significant coronal emission at 6 GHz, thus perhaps indicating a `radio-silent' population. 

It is important to note that \cite{1993ApJ...405L..63G} used X-ray data at varying energy ranges and radio data from 5 and 8 GHz to establish their relationship, while \cite{2008MNRAS.390..847L} used 5 GHz radio data and $0.2-20$ keV X-ray data to create their relationship. Meanwhile, we are using 22 GHz data and both $2-10$ keV and $14-195$ X-ray data from the \textit{Swift-XRT} and BAT telescopes. 

The $L_R$/$L_X$ relation as a function of both hard and ultra-hard X-rays and radio frequency has not been well explored until recently. \cite{2022ApJ...938...87K} and \cite{2023ApJ...952L..28R} have shown that the mm-regime and the \textit{Swift-BAT} hard and ultra-hard X-rays fluxes and luminosities correlate strongly ($1\sigma$ scatter of 0.36 and 0.22, respectively) and are a possible probe for the AGN corona. Our $L_R$/$L_X$ scatter at 22 GHz with X-rays is $2-3$ times larger than the scatters that \cite{2022ApJ...938...87K} and \cite{2023ApJ...952L..28R} report in the mm-regime. The difference in scatter supports the idea that the mm-regime arises from coronal emission that is synchrotron self-absorbed. The larger 22 GHz scatter implies that we are probing the optically thick side of the coronal emission which is contaminated by other components (i.e. free-free emission, synchrotron emission, etc.). 

\cite{2015MNRAS.454.3622B} and \cite{2022ApJS..261....7D} found a scatter of $\sim0.65$ dex when comparing the extinction corrected optical [O III] line and X-ray luminosities from AGN in the 70-month and 105-month \textit{Swift}-BAT catalog. Since [O III] is a tracer of AGN activity \citep{1981PASP...93....5B, 2002ApJS..142...35K, 2003MNRAS.346.1055K} and since we notice a similar scatter of $\sim0.5$ dex, it is likely that 22 GHz emission is as good a tracer of AGN.   

As mentioned earlier, we do see a trend with the Eddington ratio in Figure \ref{fig:22GHz-coronal-relation2} (b). Low Eddington ratio sources ($\lambda_{\text{Edd}} \leq 3$) tend to have lower X-ray and radio luminosities, and vice versa for higher Eddington ratio sources. It is important to note that this trend with the Eddington ratio weakens in flux-flux space but is still present; see Figure \ref{fig:22GHz-coronal-flux}. Most of the sources with low Eddington ratio lie above the intrinsic scatter of the data in flux-flux space.

All things considered, we find that our sample is consistent with the \cite{1993ApJ...405L..63G} correlation for coronally active stars in both the hard and ultra-hard X-rays, which hints at possible coronal emission from AGN at 22 GHz. Furthermore, we did not find any way to distinguish radio and optical morphologies along with the spectral type within the L$_R$/L$_X$ relation.

\subsection{FIR-Radio Relation} \label{subsec:FIR-Radio}

\begin{figure*}[htbp!]
    \begin{center}
    \includegraphics[trim=0mm 0mm 0mm 0mm, clip, width=2.1\columnwidth]{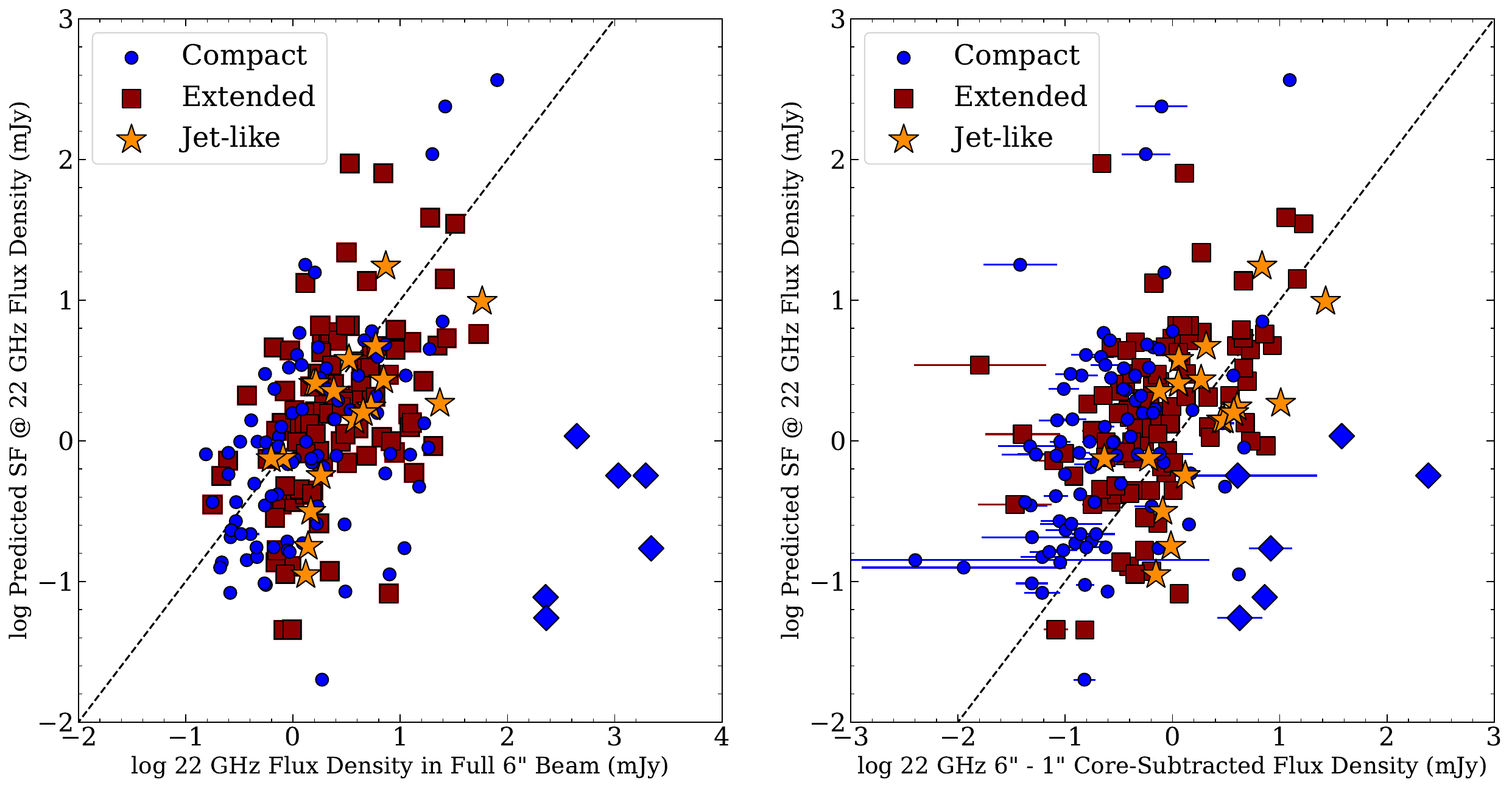}
    \caption{22 GHz observed flux densities versus the predicted 22 GHz flux densities from FIR emission associated with star formation \citep{2017MNRAS.466.3161S}, including all extended emission from the 6\arcsec~beam (left) and with the 1\arcsec ~ core subtracted out (right). Vertical error bars are shown for all points, but are often smaller than the points. The large error bars result from when the core subtraction leaves very little flux density left over. The black dashed line represents the 1-to-1 relation between the observed and predicted flux densities. Compact sources are blue circles, extended sources are green squares, and jet-like sources are orange stars, and radio-loud sources have been plotted with diamonds.}
    \label{fig:22GHz-FIR-relation}
    \end{center}
\end{figure*}

\cite{2017MNRAS.466.3161S} spectrally decomposed the \textit{Herschel} far-IR  ($8-1000$ $\mu$m) SEDs of the parent sample into integrated emission from the AGN and star formation (SF). We use the L$_{\text{FIR}}$/L$_{\text{Radio}}$ relation from \cite{1992ARA&A..30..575C} with the isolated SF component of the far-IR flux to find the predicted radio flux from star formation at 1.4 GHz:
\begin{equation}
    \text{log}\left[\frac{\text{S}_{\nu}}{\text{mJy}} \times 10^{29} \right] = \text{log} \left[ \frac{\text{FIR}}{3.75 \times 10^{12} \text{ W m}^{-2}} \right] - q
\end{equation}
where the average value of $q$ is 2.3 \citep{1992ARA&A..30..575C}, $S_{\nu}$ is the radio flux density, and FIR is the star formation emission from \cite{2017MNRAS.466.3161S}. To convert the expected 1.4 GHz flux density to 22 GHz, we must account for the thermal and non-thermal radio emission from star formation. We assume spectral indices for star formation of $0.1$ for the bremsstrahlung emission and $0.8$ for synchrotron emission and extrapolate via $S_{\nu} = \nu^{-\alpha}$ \citep[e.g.,][]{1992ARA&A..30..575C, 2018A&A...611A..55K, 2021MNRAS.507.2643A}. Using the predicted SF flux densities at 22 GHz, we then combine the non-thermal and thermal flux densities to get a total flux density that we can now compare to our observations. We use our 6\arcsec~\textit{uv}-tapered flux densities to compare to the predicted 22 GHz flux. This approximately matches the highest resolution of our \textit{Herschel} images, which is the PACS 70 $\mu$m. These images have an angular resolution 5.5\arcsec \,\citep{2014ApJ...794..152M, 2017arXiv170505693M, 2024A&A...688A.203M}, and covers roughly the same area of the galaxy as our 6\arcsec\, VLA images. We compare both the total 6\arcsec~flux density and the core subtracted (6\arcsec- 1\arcsec) flux densities to the far-IR prediction for star formation in Figure \ref{fig:22GHz-FIR-relation}. In the 6\arcsec-FIR relation (left side of Figure \ref{fig:22GHz-FIR-relation}), most of our sample is consistent with the FIR-Radio relation with a scatter of $0.56^{+0.06}_{-0.05}$ for the full sample, while the radio-loud (diamonds) sources deviate. The typical FIR-radio scatter ranges from $0.137-0.3$ dex \citep{2024MNRAS.531..708C}. Possible reasons for the higher scatter seen in our data could be from not accounting for cosmic-ray smearing, over subtraction of star formation in the 1\arcsec~beam, along with other additional uncertainties in the FIR estimates that have not been accounted for. For sources that are greater than the FIR-Radio relation, something other than star formation is providing additional radio emission at 22 GHz, such as a jet/wind or coronal component. Additionally, all of our radio-loud (diamonds) sources have higher than expected flux densities as seen in 6\arcsec~flux density comparison (see left panel of Figure \ref{fig:22GHz-FIR-relation}), implying that they likely host a genuine radio jet. In the core subtracted plot (right side of Figure \ref{fig:22GHz-FIR-relation}), over half of our sample have less than expected flux densities implying that: either the star formation prediction from \cite{2017MNRAS.466.3161S} is overestimating the star formation rate, which we consider unlikely based on the accurate FIR SED decomposition into star formation and AGN components that \cite{2017MNRAS.466.3161S} performed; or, there is significant star formation within the 1\arcsec~beam \citep{2014ApJ...781L..34M}. Having star formation within the 1\arcsec~beam is quite possible, since the beam covers a wide range of physical scales, from 35 (M~106) to 978 (LEDA 138501) parsecs, and this can be seen from the extended structures present in the $0.3\arcsec$ images. We do note a slight variation in the scatter when accounting for a volume limited sample. It is also possible that our 22 GHz observations could be overestimating the star formation rate if contributions from dust are non-negligible in some of our sources since we are at a higher frequency \citep{Murphy_2009}. \cite{2008ApJ...678..828M} has shown that a $\sim1$ kpc smearing scale for the synchrotron component is needed to match the FIR to the radio. This could account for the very mild discrepancy seen in the $6\arcsec~-1\arcsec$ difference seen in Figure \ref{fig:22GHz-FIR-relation}.

Interestingly, from Figure \ref{fig:22GHz-FIR-relation} we see that objects with extended and jet-like morphologies are the least impacted by the core subtraction; this is likely because the majority of the emission is in the extended outflow, and not the core. These sources also lie close to the \cite{1992ARA&A..30..575C} FIR-radio relation, distinct from radio-loud objects. This could imply that radio emission from the shocks along a boundary between the ISM and a radiatively driven outflow has luminosities similar to those of star formation. Alternatively, these radio morphologies could be a red herring, where star formation along bars, filaments, spiral arms, and elsewhere within the galaxy could give the appearance of outflows due to its clumpy and possibly linear structure. We do see hints of variations among the different radio morphologies, with compact (circles) sources having a dispersion of $0.62^{+0.11}_{-0.08}$ dex, extended (squares) sources having a dispersion of $0.55^{+0.08}_{-0.07}$ dex, and jet-like (stars) sources having a dispersion of $0.44^{+0.20}_{-0.12}$ dex. The scatters for the various radio morphologies remains the same in both plots of Figure \ref{fig:22GHz-FIR-relation}.

We have found within the FIR-Radio relation through 1\arcsec~core subtraction of our 6\arcsec~images that there is likely to be star formation within our 1\arcsec~beam in many of the objects. This can be seen in Figure \ref{fig:22GHz-FIR-relation}, where most of the sample has less star formation than expected after subtracting the core. This suggests that there is likely to be unresolved star formation within the central 1” in many of these galaxies. However, unresolved radio emission may also contain contributions from outflows \citep[e.g.,][]{Esquej_2013, 2013ApJ...771...63R, 2023ApJ...944...30M}. Extended and Jet-like sources are the least impacted by the subtraction, since a majority of the flux density is coming from an outflow and/or additional star formation near the core.

\subsection{Radio Detection Fraction \& Comparison to Other Frequencies} \label{subsec:Detection}

In our full sample, we only have four non-detections at 22 GHz with an rms sensitivity of $\sim 16$ $\mu$Jy, yielding a detection fraction of $98 \pm 1\%$ (227/231) from \textit{Swift}-BAT selected sources. This result is similar to what Paper II found in the previous phase of this radio survey. In almost all detected sources, a compact core is present. The core was not dominant for two jet-like sources (II Zw 83 and Fairall 272); this could result from the fact that the jet lobes are brighter than the core, along with the fact that with our 1\arcsec~beam the lobes could blend with the radio core or indicate that these sources are dual AGN. 

To better understand the non-detections in our sample and how our 22 GHz survey compares to other radio surveys, we cross-matched our sample with archival radio surveys at other lower frequencies and resolutions using the CIRADA Image Cutout Web Service\footnote{\url{http://cutouts.cirada.ca/}}.Given a typical AGN synchrotron spectral slope of $-0.6$, we expect BAT sources to be more luminous at lower frequencies. Table \ref{tab:detection_fraction} shows the fraction of our sample detected by each survey. To not penalize any of the surveys for having footprints in smaller and disparate regions of the sky, we only considered sources from our sample within each survey footprint. Table \ref{tab:detection_fraction} contains the frequency range, resolution, physical scale probed, detection rate, and sensitivity of each all-sky radio survey that we have examined in this work. 

\begin{deluxetable*}{cccccccc}
\tablecaption{Comparison of our VLA 22 GHz radio survey to other all-sky radio surveys.}
\tablehead{
\colhead{Survey} & \colhead{Frequency Range} & \colhead{Resolution} & \colhead{Physical Scale} & \colhead{Sensitivity} & \colhead{Number of} & \colhead{Detection} & \colhead{Reference} \\  Name & MHz & \arcsec & pc & mJy beam$^{-1}$ & Sources & Rate &}
\colnumbers
\startdata
GLEAM    & $72-231$       & $\sim120$ & $4200-117360$ & $\sim11.3$ & $\sim57$ & $\sim37\%$ & \cite{2015PASA...32...25W}\\
         &                &           &               &            &          &            & \cite{2017MNRAS.464.1146H}\\
LoTSS    & $120-168$      & $6$       & $210-5868$    & $0.083$    & $56$     & $100\%$    & \cite{Shimwell_2019}      \\
VLITE    & $236-492$      & $\sim6$       & $350-9780$    & $\sim3$    & $77$     & $62\%$     & \cite{2016arXiv160303080C, 2016SPIE.9906E..5BC}\\
RACS-low & $745-1033$     & $25$      & $875-24450$   & $0.4$      & $148$    & $94\%$     & \cite{2020PASA...37...48M}\\
         &                &           &               &            &          &            & \cite{2021PASA...38...58H}\\
NVSS     & $1304-1477$    & $45$      & $1575-44010$  & $0.45$     & $208$    & $82\%$     & \cite{1996IAUS..175..503C}\\
         &                &           &               &            &          &            & \cite{1998AJ....115.1693C}\\
FIRST    & $1343-1457$    & $5$       & $175-4890$    & $0.15$     & $92$     & $84\%$     & \cite{1995ApJ...450..559B}                     \\
VLASS    & $1000-5000$    & $\sim2.5$ & $87.5-2445$   & $0.07$     & $233$    & $90\%$     & \cite{2020PASP..132c5001L}                     \\
Our Work & $18000-26000$  & $1$       & $35-978$      & $0.016$    & $231$    & $98\%$     & Paper I, Paper II, and this work               \\
\enddata
\label{tab:detection_fraction}
\tablecomments{Detection rates of our \textit{Swift}-BAT detected sources from various radio surveys at different frequencies and resolutions. Columns are: (1) Survey name, (2) Frequency range that the survey covered, (3) Resolution in arcseconds, (4) Physical scale in parsecs that the surveys probed, (5) Sensitivity of the survey in mJy $\text{beam}^{-1}$, (6) Number of sources from our original 22 GHz survey that could have been in each survey, (7) Detection rate of each survey, and (8) Survey reference(s).}
\end{deluxetable*}

\subsubsection{Comparison to All-sky Radio Surveys}\label{subsubsec:all-sky}

In general, all the sensitive radio surveys \citep[i.e., excluding GLEAM and VLITE;][]{2015PASA...32...25W, 2017MNRAS.464.1146H, 2016arXiv160303080C, 2016SPIE.9906E..5BC} exhibit high detection fractions ($>80\%$) for our sample that is available to the respective survey's footprint. The detection fractions of GLEAM and VLITE are substantially lower than those of other surveys, because their sensitivity threshold is above our median flux density of 0.928 mJy. However, one must consider that many of these detections are most likely contaminated by extended star formation due to the larger beam sizes, creating additional problems for interpretation of contributing components.

Overall, when comparing our survey to other all-sky radio surveys, we find that our high sensitivity at 22 GHz yields a very high detection fraction that other all-sky surveys roughly match. It can be seen that LoTSS and GLEAM both have only 56 sources within their survey footprint, but the difference in survey sensitivity yields two different detection rates.

\subsubsection{Comparison to Targeted Radio Surveys}
\label{subsubsec:related}

In addition to all the all-sky surveys that have taken place over the years, several targeted surveys have also taken place. In particular, several targeted high-frequency ($\sim$ 22 GHz) radio surveys of X-ray selected AGN have taken place. In this section, we will explore how our complete survey compares to these targeted surveys.

\cite{2013MNRAS.431.2471B} completed a very similar X-ray selected radio survey to ours with the Australia Telescope 20 GHz (AT20G) Survey \citep{2010MNRAS.402.2403M} and the 60-month all sky \textit{Swift}-BAT survey \citep{2012ApJ...749...21A}. They found 37 BAT AGN with radio counterparts in the AT20G. The AT20G survey has an rms sensitivity of $\sim 10$ mJy with a resolution ranging from $\sim 10-30\arcsec$. This sensitivity suggests that AT20G would only be able to detect 32 of our sources (those with flux density $\ge 10$ mJy), approximately 14\% ($\sim32/231$) of our sample, which is consistent with the 37 detected radio sources reported by \cite{2013MNRAS.431.2471B}. Note the 14\% of our sample presented above does not take into account the declination limits of the AT20G survey. It is better to view the above percentage as if AT20G had taken place at the same location as VLA.


\cite{2020MNRAS.495.3943C} investigated a hard X-ray selected sample from \textit{Integral}, finding a detection rate of 15/16 of their sources at 22 and 45 GHz from the VLA in C-configuration. Their 22 GHz data has an rms sensitivity of $\sim 21$ $\mu$Jy, and their 45 GHz data has an rms sensitivity of $\sim 47$ $\mu$Jy. This survey reproduces a very similar detection rate to ours, although it only has 16 sources. In addition, their survey is the only survey so far that has VLA 22 GHz follow-up of X-ray detections, and they selected the C-configuration to get 1\arcsec ~ resolution, which maps to kiloparsec-scale or smaller structures in low-redshift ($z \leq 0.323$) galaxies. \cite{2022MNRAS.515..473P} followed up \cite{2020MNRAS.495.3943C} with a wide-band radio study in the 5-45 GHz regime. They matched the 1\arcsec ~ resolution at 22 GHz with the VLA in C-configuration with their lower frequencies (5-15 GHz; C (5 GHz), X (10 GHz), and Ku (15 GHz) bands) along with matching the high sensitivity of \cite{2020MNRAS.495.3943C}. They find a very high detection fraction across all bands. They detected all the sources for which they had observations at 5, 10, and 15 GHz. At 22 GHz 26/29 sources, and at 45 GHz they detect 24/29 sources. Their high detection rates from 5-45 GHz are a result of their high sensitivities, just like our sample. If we apply the results of \cite{2022MNRAS.515..473P} to our sample, we should expect a uniform detection fraction down to lower frequencies. This is supported by the LoTSS detection rate, which is the only low-frequency survey that is close to our sensitivity level. 

The Fundamental Reference AGN Monitoring Experiment (FRAMEx) is a unique VLBA observing experiment that is designed to study the apparent positions and morphologies of AGN at extremely high resolution \citep{2020jsrs.conf..165D}. \cite{2021ApJ...906...88F} (FRAMEx I) conducted a survey very similar to ours, but with two key differences: 1. They observed with the VLBA rather than VLA, achieving milliarcsecond resolution instead of arcsecond resolution. 2. They carried out contemporaneous observations with the \textit{Swift}-BAT and C-band (5 GHz) on the VLBA with an rms sensitivity of $\sim 20$ $\mu$Jy beam$^{-1}$, similar to ours. Thus, one would expect around the same detection rate as our survey since FRAMEx has a high sensitivity paired with the high-resolution power of VLBA. However, among the sample of 25 \textit{Swift}-BAT selected AGN investigated in \cite{2021ApJ...906...88F}, they only had 9 detections showing compact, extended, and binconical structures, yielding a 36\% detection rate. Surprisingly, 19/25 of their sources had archival VLA C-band detections. By using the Black Hole Fundamental Plane to compare their VLA and VLBA data, they found that the archival C-band data are consistent with the fundamental plane (i.e., it has the radio luminosity consistent with the scale-invariant jet model using the measured black hole mass and X-ray luminosity), the VLBA data were strongly discrepant, especially the non-detections. They argued that the discrepancy between the VLA and VLBA scales could be due to extranuclear radio emission, such as an AGN jet/wind interacting with the ISM. \cite{2022ApJ...936...76S} (FRAMEx III) expanded on the original sample of \cite{2021ApJ...906...88F} with 9 new sources and deeper observations of 9 original sources at 5 GHz. These VLBA observations had an rms sensitivity of $\sim 8$ $\mu$Jy; they recovered three original sources and detected two new sources, bringing their detection rate to 41\% (14/34 detections). Thus, the plane may be less fundamental than previously thought. \cite{2024ApJ...961..109S} (FRAMEx V) expanded to include L-band (1.6 GHz), X-band (8.6 GHz), and K-band along with C-band observations of 12 sources from their original sample presented in FRAMEx I. They detected all of the sources at L-band and X-band, 9/12 of the sources at C-band, and 6/12 sources at K-band. Similarly, \cite{2019MNRAS.488.4317B} found similar results using the Korean VLBI Network (KVN) and VLBA at 22 GHz with a detection rate of 21\% (10/47 detections) at 22 GHz and 0 detections out of 95 BAT AGN sources with a blind fringe survey at 15 GHz. Since \cite{2019MNRAS.488.4317B} and FRAMEx V probes a physical scale of $0.0345 - 0.372$ parsecs at 22 GHz with VLBA and our VLA survey covers $35-978$ parsecs, then the source of high frequency emission may, therefore, occupy an intermediate scale between VLBA and VLA scales.

Comparing our 22 GHz VLA survey with the others mentioned above, one can see that our survey sensitivity stands out as one of the highest sensitivities. Additionally, as a result of this high sensitivity, our high detection is among the highest when compared to other high-frequency surveys. \cite{2020MNRAS.495.3943C} achieves similar results to us at 22 GHz, but with a much smaller sample. Meanwhile, the VLBA studies at 22 GHz (\cite{2019MNRAS.488.4317B} and FRAMEx V) fail to detect more than 50\% of their sample. In general, our survey is at present the most sensitive high-frequency radio survey of a large sample of X-ray selected AGN.

\subsection{Non-detections} \label{subsec:Non-detection}

As mentioned in Section \ref{sec:sample}, we only have four non-detections in our sample: Mrk\,352, a narrow-line Seyfert 1, which is known to be variable in the X-rays \citep[e.g.,][]{2006AJ....132..321D, 2016A&A...592A..74S}, Mrk\,653, IRAS\,03219+4031, and MCG-02-02-095. Narrow-line Seyfert 1 galaxies are also known to vary rapidly in the radio \citep[e.g.,][]{2024MNRAS.532.3069J}. If we consider the 12-month gap between the 58-month and 70-month \textit{Swift}-BAT survey along with our non-simultaneous radio observations during this time, then this would imply that these four sources may be highly variable in the radio and/or the X-ray. This is supported by the variability shown in the X-ray light curves provided by the \textit{Swift}-BAT 105-month hard X-ray survey \citep{2018ApJS..235....4O} \footnote{\url{swift.gsfc.nasa.gov/results/bs105mon/}}. 

Of the cross-matched surveys mentioned above, only two non-detections in our survey are detected. Mrk 352 is detected with LoTSS and VLASS, and IRAS\,03219+4031 is detected with NVSS and VLASS. In addition to this, we searched NASA/IPAC Extragalactic Database (NED) \citep{https://doi.org/10.26132/ned1}\footnote{\url{http://ned.ipac.caltech.edu/}} and the literature more broadly; no archival radio detections of the other sources were found.


\begin{figure*}[htbp!]
    \includegraphics[trim=0mm 0mm 0mm 0mm, clip, width=2.1\columnwidth]{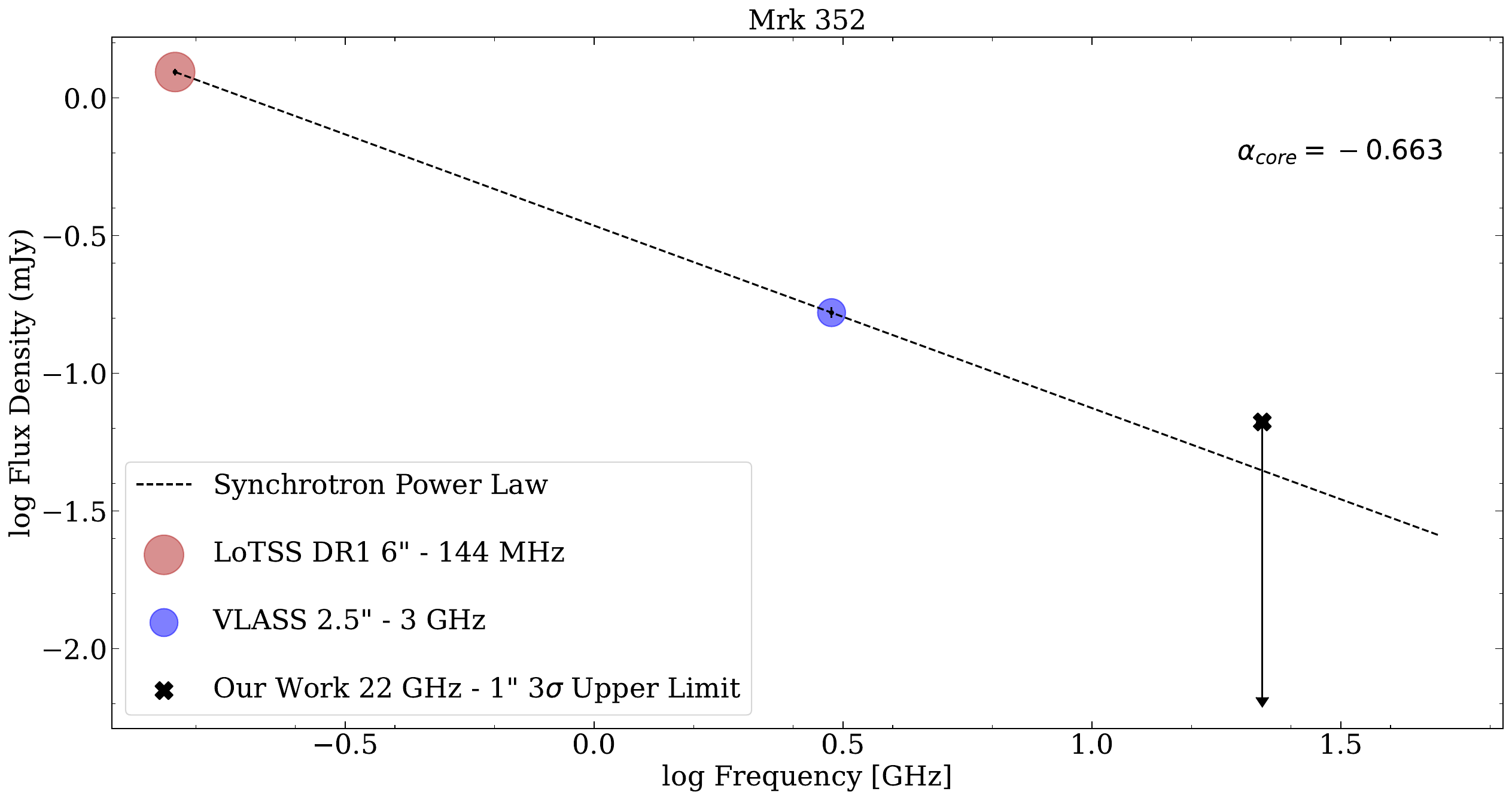} \\
    \quad
    \includegraphics[trim=0mm 0mm 0mm 0mm, clip, width=2.1\columnwidth]{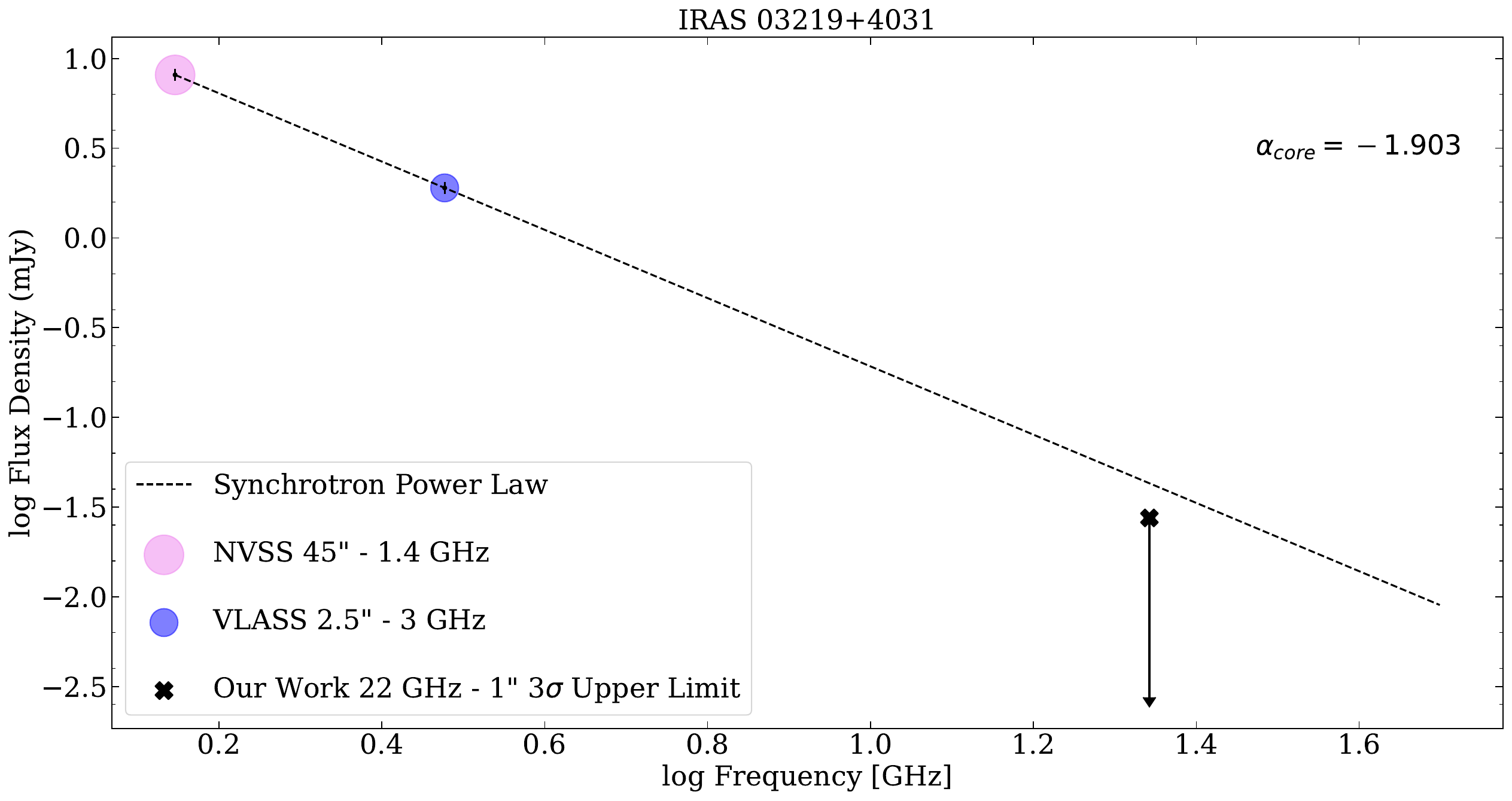} \\
  \caption{Radio SED of Mrk\,352 and IRAS\,03219+4031. The data included in these radio SEDs are from LoTSS DR1 \citep[red;][]{Shimwell_2019}, VLASS \citep[blue;][]{2020PASP..132c5001L}), NVSS \citep[violet;][]{1996IAUS..175..503C, 1998AJ....115.1693C} and our 3$\sigma$ upper limit. The linear regression represents a simple synchrotron power law with spectral index as printed in the top right of the plot. $\alpha_{\text{core}}$ represents the spectral index derived from a single synchrotron power-law.}
  \label{fig:non-detections}
\end{figure*}

Since Mrk 352 and IRAS 03219+4031 are the only sources that had multiple detections at different frequencies, we constructed their radio spectral energy distribution (SED) to see where our 22 GHz 3$\sigma$ non-detection falls in relation to a simple synchrotron power-law, shown in Figure \ref{fig:non-detections}. The lower frequency data points were fit with a simple linear regression under the assumption that Mrk 352 and IRAS 03219+4031 are entirely made up of synchrotron emission in the radio regime. From linear regression, we find a non-thermal spectral index of -0.663 for Mrk 352, which is consistent with typical AGN \citep[e.g.,][]{2005AJ....130.1358N, 2010MNRAS.409..541S}. Our 22 GHz upper limit falls slightly above the extrapolation of this synchrotron power-law, implying that the radio emission in this galaxy deviates from a single, non-thermal, power law seen at lower frequencies. Instead, our upper limit implies an even steeper non thermal spectral index, possibly indicating that the synchrotron emission has started to ``age'' out due to the AGN turning off \citep[e.g.][]{1987MNRAS.225....1A, 1990ApJ...353..476W}. For IRAS\,03219+4031, we found a non-thermal spectral index of -1.903, which is much steeper than Mrk\,352, and the steepness of this spectral index indicates that the synchrotron emission has ``aged''. This spectral index should be treated as a lower limit because of the vast difference in resolution between NVSS and VLASS. Additionally, the NVSS flux density is likely to be enhanced because nearby galaxies (2MASS\,J03251496+4041471 and IGR\,J03249+4041) could be contributing to the radio emission. A higher resolution 1.4 GHz data point would help constrain the spectral index of this source. 

Extending this result to the other three radio non-detections that did not have any ancillary radio detections implies that there may be a radio population that is anomalously quiet and most likely will not be detected until future observations with the next generation VLA (ngVLA) and the Square Kilometre Array (SKA) are conducted \citep{2018ApJ...859...23N}.

\subsection{Serendipitous Radio Sources} \label{subsec:Serendipitous}

In addition to the target sources, our survey images included a total of 29 serendipitous radio sources in the same field of view. All of these sources are listed in Table \ref{tab:rbongs} along with their flux densities. In $37.9 \pm 9\%$ (11/29) of the cases, these are associated with optical galaxies identified from the DESI Legacy Imaging Survey \citep{2019AJ....157..168D}\footnote{\url{https://www.legacysurvey.org/viewer/}}, NED and Pan-STARRS $1$ g-band imaging \citep{2020ApJS..251....7F}. Most of these optically identified sources have a redshift similar to that of our original science target. In $13.8 \pm 6\%$ (4/29) of the cases, serendipitous sources are likely to be associated with the host galaxy, such as the ultraluminous X-ray source (ULX) CXO J133815.6+043255 in NGC 5252 \citep{2017ApJ...844L..21K, 2023ApJ...956....3S} and [HU2001] J131330.1+363537 in NGC 5033 \citep{2007MNRAS.379..275P}; the latter was first observed by \cite{2001ApJS..133...77H} in the radio, and later found to have a faint Chandra counterpart in \cite{2019ApJS..243....3L}. We were able to recover NGC 232 in our \textit{uv}-tapered 3"~and 6"~images. \cite{2021ApJS..257...61Y} classified this source as star-forming based on the \textit{Chandra} spectrum. Since NGC 232 has been classified as star-forming in the soft X-rays ($<2$ keV), it makes sense that this source is not in the \textit{Swift}-BAT catalog since \textit{Swift} probes higher energy X-rays that usually come from the AGN. 

In 48.3\% (14/29) of cases, no optical counterpart is found to the aforementioned depths, and these are most likely distant radio galaxies (e.g., PMN J2001-1808). We classify these sources as optically faint radio galaxies. Table \ref{tab:rbongs} lists the flux densities we have measured for all serendipitous sources at 22 GHz in both our C- and B-array data. Additionally, we have noted the AGN's name that has the optically faint radio galaxy in its image (i.e. VLA CGCG468-002). 

Most of these optically faint radio galaxies are at a distance of $\sim$ 20\arcsec~or further away from our main target source. The exception is the radio galaxy near ESO 553-G043, which is at a distance of $\sim$ 9\arcsec. Compared to a few lower frequency surveys and accounting for their footprints, we find the following: VLASS detects 86\% (25/29), FIRST detects 50\% (6/12), and LoTSS detects 71\% (5/7) of these optically faint radio galaxies. 

\begin{deluxetable*}{c c c c c c c c c c c}
\tabletypesize{\scriptsize}
\tablecaption{22 GHz Serendipitous Radio Sources \label{Serendipitous}}
\tablehead{
\colhead{Name} & \colhead{RA} & \colhead{Dec} & \colhead{0.\arcsec3} & \colhead{S$_{\nu, 0.\arcsec3}$} & \colhead{1\arcsec} & \colhead{S$_{\nu, 1\arcsec}$}  & \colhead{3\arcsec} & \colhead{S$_{\nu, 3\arcsec}$}  & \colhead{6\arcsec} & \colhead{S$_{\nu, 6\arcsec}$} \\  & &  & Morph. & mJy  & Morph. & mJy & Morph. & mJy& Morph. & mJy}
\colnumbers
\startdata 
NGC 232                  & 00:42:45.8 & -23.33.40.8 & \nodata & \nodata $\pm$ \nodata & \nodata & \nodata $\pm$ \nodata & 2       & 0.253   $\pm$ 0.013   & 1 & 0.361   $\pm$ 0.013          \\
LEDA 212995              & 02:28:14.3 & +31.19.00.3 & \nodata & \nodata $\pm$ \nodata & 1       & 0.351   $\pm$ 0.020   & 1       & 0.357   $\pm$ 0.011   & 1 & 0.601   $\pm$ 0.017          \\
ZEH2003 RX J0241.5+0711 4 & 02:41:34.2 & +07.10.51.5 & \nodata & \nodata $\pm$ \nodata & 1       & 0.221   $\pm$ 0.004   & 1       & 0.403   $\pm$ 0.011   & 1 & 0.793   $\pm$ 0.013          \\
NVSS J035406+024915   & 03:54:06.6 & +02.49.14.2 & \nodata & \nodata $\pm$ \nodata & 1       & 0.690   $\pm$ 0.004   & 1       & 0.692   $\pm$ 0.007   & 1 & 0.726   $\pm$ 0.004          \\
2MFGC 4194               & 05:08:21.2 & +17.22.08.3 & \nodata & \nodata $\pm$ \nodata & 2       & 3.015   $\pm$ 0.041   & 2       & 4.525   $\pm$ 0.026   & 1 & 4.651   $\pm$ 0.015          \\
VLA CGCG468-002        & 05:08:23.6 & +17.21.02.7 & \nodata & \nodata $\pm$ \nodata & 1       & 1.602   $\pm$ 0.006   & 1       & 1.728   $\pm$ 0.010   & 1 & 1.749   $\pm$ 0.013          \\
2MASX  J06021107+2828382     & 06:02:11.0 & +28.28.37.9 & \nodata & \nodata $\pm$ \nodata & 1       & 0.191   $\pm$ 0.015   & 1       & 0.192   $\pm$ 0.022   & 1 & 0.258   $\pm$ 0.010          \\
(2017)    & 06:02:11.0 & +28.28.37.9& \nodata & \nodata $\pm$ \nodata & 1       & 0.336   $\pm$ 0.007   & 1       & 0.334   $\pm$ 0.015   & 1 & 0.390   $\pm$ 0.013          \\
VLA UGC03995A          & 07:44:07.1 & +29.15.56.3 & \nodata & \nodata $\pm$ \nodata & 2       & 0.871   $\pm$ 0.012   & 1       & 0.961   $\pm$ 0.007   & 1 & 1.060   $\pm$ 0.013          \\
2MASS J09191594+5526559        & 09:19:16.0 & +55.26.56.4 & \nodata & \nodata $\pm$ \nodata & 2       & 0.641   $\pm$ 0.02    & 1       & 0.696   $\pm$ 0.013   & 1 & 0.732   $\pm$ 0.018          \\
WISEA J094204.48+234042.8        & 09:42:04.5 & +23.40.42.3 & 1       & 0.106   $\pm$ 0.003   & 2       & 0.141   $\pm$ 0.006   & 1       & 0.220   $\pm$ 0.007   & 1 & 0.454   $\pm$ 0.013          \\
6C 110414+724854              & 11:07:41.7 & +72.32.36.1 & \nodata & \nodata $\pm$ \nodata & \nodata & \nodata $\pm$ \nodata & \nodata & \nodata $\pm$ \nodata & 1 & 1.900   $\pm$ 0.120          \\
NVSS J114540-182838 (1)  & 11:45:40.9 & -18.28.41.0 & \nodata & \nodata $\pm$ \nodata & 2       & 0.248   $\pm$ 0.005   & 1       & 0.325   $\pm$ 0.006   & 1 & 0.419   $\pm$ 0.005          \\
NVSS J114540-182838 (2)  & 11:45:39.4 & -18.28.26.8 & \nodata & \nodata $\pm$ \nodata & 1       & 0.097   $\pm$ 0.003   & 1       & 0.262   $\pm$ 0.008   & 1 & 0.364   $\pm$ 0.015          \\
ICRF J120913.6+433920            & 12:09:13.4 & +43.39.17.0 & \nodata & \nodata $\pm$ \nodata & \nodata & \nodata $\pm$ \nodata & 2       & 2.089   $\pm$ 0.068   & 2 & 2.800   $\pm$ 0.130          \\
FIRST J124143.7+350404           & 12:41:43.7 & +35.04.05.8 & \nodata & \nodata $\pm$ \nodata & 1       & 0.504   $\pm$ 0.017   & 1       & 2.235   $\pm$ 0.046   & 1 & 0.618   $\pm$ 0.010          \\
HU2001 J131330.1+363537         & 13:13:30.1 & +36.35.37.1 & \nodata & \nodata $\pm$ \nodata & 1       & 0.325   $\pm$ 0.018   & 1       & 0.352   $\pm$ 0.003   & 1 & 0.385   $\pm$ 0.008          \\
WISEA J133545.08+030022.6           & 13:35:45.0 & +03.00.22.6 & \nodata & \nodata $\pm$ \nodata & 1       & 0.455   $\pm$ 0.059   & 1       & 0.536   $\pm$ 0.103   & 1 & 0.645   $\pm$ 0.167          \\
CXO J133815.6+043255              & 13:38:15.6 & +04.32.55.3 & \nodata & \nodata $\pm$ \nodata & 1       & 0.527   $\pm$ 0.024   & 1       & 0.588   $\pm$ 0.004   & 1 & 0.591   $\pm$ 0.005          \\
WISEA J135133.17-181257.4     & 13:51:33.1 & -18.12.57.0 & \nodata & \nodata $\pm$ \nodata & 1       & 0.257   $\pm$ 0.007   & 1       & 0.272   $\pm$ 0.004   & 2 & 0.297   $\pm$ 0.069          \\
VLA Mrk 279            & 13:52:50.5 & +69.18.22.4 & 0       & \nodata $\pm$ \nodata & 0       & \nodata $\pm$ \nodata & 1       & 0.278   $\pm$ 0.061   & 1 & 0.292   $\pm$ 0.011          \\
MCG+12-13-024            & 13:53:11.8 & +69.18.41.6 & 0       & \nodata $\pm$ \nodata & 0       & \nodata $\pm$ \nodata & 1       & 0.224   $\pm$ 0.061   & 1 & 0.282   $\pm$ 0.054          \\
SSTSL2 J150646.23+035121.5     & 15:06:46.2 & +03.51.21.4 & \nodata & \nodata $\pm$ \nodata & 1       & 0.340   $\pm$ 0.023   & 1       & 0.389   $\pm$ 0.007   & 1 & 0.442   $\pm$ 0.009          \\
NGC 6230                 & 16:50:46.7 & +04.36.16.7 & \nodata & \nodata $\pm$ \nodata & 1       & 8.141   $\pm$ 0.023   & 1       & 8.306   $\pm$ 0.056   & 1 & 8.418   $\pm$ 0.074          \\
WISEA J174316.48+624900.4       & 17:43:16.5 & +62.49.00.1 & \nodata & \nodata $\pm$ \nodata & 1       & 0.194   $\pm$ 0.016   & 1       & 0.239   $\pm$ 0.016   & 1 & 0.282   $\pm$ 0.023          \\
MCG+07-37-031     & 18:16:11.7 & +42.39.37.0 & 2       & 4.496   $\pm$ 0.006   & 2       & 6.656   $\pm$ 0.044   & 1       & 7.880   $\pm$ 0.056   & 1 & 7.931   $\pm$ 0.026          \\
SSTSL2 J193734.59-061402.2      & 19:37:34.6 & -06.14.02.3 & \nodata & \nodata $\pm$ \nodata & 0       & \nodata $\pm$ \nodata & 1       & 0.321   $\pm$ 0.043   & 1 & 0.329   $\pm$ 0.011          \\
PMN J2001-1808             & 20:00:58.9 & -18.08.57.5 & \nodata & \nodata $\pm$ \nodata & 2       & 0.260   $\pm$ 0.003   & 2       & 0.542   $\pm$ 0.023   & 2 & \nodata $\pm$ \nodata        \\
NGC 6921                 & 20:28:28.9 & +25.43.24.3 & \nodata & \nodata $\pm$ \nodata & 1       & 0.225   $\pm$ 0.009   & 1       & 0.354   $\pm$ 0.011   & 1 & 0.440   $\pm$ 0.023          \\
HU2001 J230458.7+121917          & 23:04:58.5 & +12.19.19.1 & \nodata & \nodata $\pm$ \nodata & 2       & 0.208   $\pm$ 0.014   & 1       & 0.368   $\pm$ 0.013   & 1 & 0.370   $\pm$ 0.017          \\
VLA ESO 553-G043       & 05:26:27.9 & -21.17.12.4 & \nodata & \nodata $\pm$ \nodata & 2       & 0.107   $\pm$ 0.005   & 2       & 0.419   $\pm$ 0.007   & 2 & 0.488   $\pm$ 0.013          \\
\enddata
\tablecomments{22 GHz Serendipitous Radio Sources. Columns are (1) Name, (2) Right Ascension, (3) Declination, (4, 6, 8, 10) are the respective morphology class at those resolutions, and (5, 7, 9, 11) are the respective flux densities at those resolutions in mJy. The morphology class is described in detail in Section \ref{sec:morphologies}.}
\label{tab:rbongs}
\end{deluxetable*}

\subsection{Discussion} \label{subsec:Discussion}

The most not so striking result after the previous discussion of our survey is our 98\% detection fraction at 22~GHz. \cite{2018MNRAS.478..399B} has shown that NGC 2992, which has a normal power-law SED from $1.4-15$~GHz, exhibits a strong 22~GHz spectral excess above an extrapolation of that power law. Their 100 GHz data point is above the synchrotron power law as well. Ultimately, the 22 and 100 GHz data could represent a spectral component originating in the corona rather than synchrotron jet emission. We interpret our 22 GHz survey of BAT AGN in terms of potential constraints on AGN coronae. As an aside, if we assume that the rest of the sample has a similar 22 GHz spectral excess that deviates from a synchrotron power-law, then it would be fairly reasonable to assume that we should detect almost all of our sample. 

In Figures \ref{fig:22GHz-coronal-relation2} (b) and \ref{fig:22GHz-coronal-flux}, we observe a trend with $\lambda_\mathrm{Edd}$ in the luminosity-luminosity and flux-flux plots, such that higher $\lambda_\mathrm{Edd}$ are at higher luminosities and fluxes. Lower $\lambda_\mathrm{Edd}$ objects may be less likely to follow the coronal relation, as these objects are more likely to launch jets, and, within our sample, these sources have higher radio fluxes than the rest of the sample. \cite{2019MNRAS.482.5513L} found that radio-quiet quasars with $\lambda_\mathrm{Edd}$ $> 0.3$ tend to show compact optically thin synchrotron components, while sources with lower $\lambda_\mathrm{Edd}$ exhibit optically thick synchrotron components.

More recently, \cite{2023ApJ...952L..28R} used high spatial-resolution observations from the Atacama Large Millimeter/submillimeter Array (ALMA) to probe the nuclear region of 26 \textit{Swift}-BAT selected AGN. They found a tight correlation between their 100 GHz fluxes and the intrinsic, absorption-corrected X-ray emission, possibly hinting at AGN coronal activity as the origin of both wavebands. If this is true, then the AGN's corona, if strong enough, could be prevalent at lower frequencies, such as 22 GHz. More detailed spectral modeling across a wide frequency range is required before confirming this additional spectral component. But, as mentioned earlier, the scatter at 22 GHz for $L_R$/$L_X$ is much higher than what is reported at 100 GHz, and this increase in the scatter could be due to other sources of radio emission (i.e., free-free, synchrotron, etc.). Additionally, at 22 GHz we are likely probing higher optical depths on the self-absorbed regime of synchrotron coronal emission, which will increase our scatter.

\section{Conclusion} \label{sec:Conclusion}

Our 22 GHz 1\arcsec~resolution imaging campaign of 231 \textit{Swift}-BAT selected sources has been presented above. Papers I and II formed the first two phases of our campaign, and we have expanded on it by adding 132 additional sources. Similarly to the previous phases, the observed morphologies of the entire sample consisted of compact/unresolved, extended, and jet-like or biconical morophologies. The 1$\arcsec$ core radio emission encompasses spatial regions ranging from 35 to 978 parsecs, and most of the extended emission and biconical outflows are on the order of kiloparsec scales. In this paper, we focus on the core radio emission and compare it to hard and ultra-hard X-ray fluxes and luminosities along with predictions from the FIR-Radio correlation. Our conclusions are as follows: 
\begin{enumerate}

\item Of our full sample of 231 AGN, we only have four AGN non-detections at 22 GHz, a $98\pm1$\% (227/231) detection fraction, which seems to result from our high sensitivity. Lower frequency surveys that match our sensitivity have a similar detection fraction. 

\item Similarly to Papers I and II, we find that our sample agrees with the \cite{1993ApJ...405L..63G} correlation for coronally active stars in the hard X-rays ($\text{L}_R/\text{L}_{2-10 \text{keV}} \sim 10^{-5}$) and ultra-hard ($\text{L}_R/\text{L}_{14-195 \text{keV}} \sim 10^{-5.5}$). Thus, hinting at possible coronal emission from AGNs at 22 GHz. 

\item We did not find any way to distinguish extended star-forming regions from biconical outflows through the L$_{R}$/L$_{X}$ relation. Nor did we find any distinction when including the optical, spectral, or radio morphology within the L$_{R}$/L$_{X}$ relation. 

\item Through 1\arcsec~core subtraction of our 6\arcsec~images and the FIR-Radio relation, we have found that there is likely to be star formation within our 1\arcsec~beam in many of our objects, especially those with compact morphologies. We also found that the extended and jet-like morphologies are the least impacted by the core subtraction, since the majority of their flux density is within the outflow. 
\end{enumerate}

The ngVLA and SKA are at the forefront of the next generation radio interferometers. The unprecedented resolution ($\sim0.26\arcsec$) and sensitivity ($\sim0.17 \mu\,\text{Jy}$) of these next generation interferometers will allow us to trace AGN with X-ray emission similar to that discussed in this paper ($z \leq 0.05$) to more distant AGN at the peak of black growth ($z = 1-2$) and beyond. Since we have detected almost all of our sources with a sensitivity of $16\, \mu\text{Jy}$, the ngVLA and SKA, with a sensitivity of $\sim100$ times higher, could allow us to probe deeper than our original survey.

This 22~GHz radio atlas, both flux data and imaging products, is public and available to the wider astronomy community (see Section \ref{subsec:sample}), and represents the largest high-frequency, high-resolution imaging survey of AGN to date. As additional low-frequency all-sky radio surveys proliferate in the upcoming years, we hope these data will provide an important and complementary database of flux densities and morphologies. Additionally, these data will be of great use in better understanding the inner mechanisms of the radio regime for AGN through future radio SED work. 

\section{Acknowledgments} \label{sec:Acknowledgments}
The authors are very thankful for the referee's valuable comments. We would like to thank the NRAO helpdesk for the helpful feedback they have provided with our data reduction and other questions we have had. 
This research used APLpy, an open source plotting package for Python \citep{aplpy2012, aplpy2019}. This research has used the CIRADA cutout service at \url{cutouts.cirada.ca}, operated by the Canadian Initiative for Radio Astronomy Data Analysis (CIRADA). CIRADA is funded by a grant from the Canada Foundation for Innovation 2017 Innovation Fund (Project 35999), as well as by the Provinces of Ontario, British Columbia, Alberta, Manitoba and Quebec, in collaboration with the National Research Council of Canada, the US National Radio Astronomy Observatory and Australia’s Commonwealth Scientific and Industrial Research Organisation. This research has used the SIMBAD database, operated at CDS, Strasbourg, France. The Pan-STARRS1 Surveys (PS1) and the PS1 public science archive have been made possible through contributions by the Institute for Astronomy, the University of Hawaii, the Pan-STARRS Project Office, the Max-Planck Society and its participating institutes, the Max Planck Institute for Astronomy, Heidelberg and the Max Planck Institute for Extraterrestrial Physics, Garching, The Johns Hopkins University, Durham University, the University of Edinburgh, the Queen's University Belfast, the Harvard-Smithsonian Center for Astrophysics, the Las Cumbres Observatory Global Telescope Network Incorporated, the National Central University of Taiwan, the Space Telescope Science Institute, the National Aeronautics and Space Administration under Grant No. NNX08AR22G issued through the Planetary Science Division of the NASA Science Mission Directorate, the National Science Foundation Grant No. AST-1238877, the University of Maryland, Eotvos Lorand University (ELTE), the Los Alamos National Laboratory, and the Gordon and Betty Moore Foundation. MM thanks Jonelle Walsh, Refa Al-Amri, Ryne Dingler, and Divya Mishra for helpful feedback on MCMC modeling and help with plotting in Python and LaTex. KLS gratefully acknowledges helpful discussions with Travis Fischer, Nathan Secrest, and Megan Johnson. KLS and MM would also like to thank Tracy Clarke, Wendy Peters, and the Naval Research Lab for providing us access to the VLITE reduced data for our sources. We gratefully acknowledge the funding support from ANID in the form of the CATA-BASAL project FB210003 (CR, FEB); the Millennium Science Initiative Program ICN12\_009 (FEB); and FONDECYT Regular grants \#1230345 (CR) and \#1241005 (FEB). KO acknowledges support from the Korea Astronomy and Space Science Institute  under the R\&D program supervised by the Ministry of Science and ICT, Project \#2024-1-831-01, and the National Research Foundation of Korea (NRF-2020R1C1C1005462).

%

\vspace{5mm}
\facilities{\textit{Herschel}, \textit{Swift}-BAT, VLA}


\software{APLpy \citep{aplpy2012, aplpy2019}, Astropy \citep{astropy:2013, astropy:2018, astropy:2022}, CARTA \citep{2020AAS...23536411O}, CASA \citep{2007ASPC..376..127M, TheCASAteam_2022}, dynesty \citep{2020MNRAS.493.3132S, sergey_koposov}, Matplotlib \citep{4160265}, NumPy \citep{2020Natur.585..357H}, pandas \citep{mckinney-proc-scipy-2010, reback2020pandas}, SciPy \citep{scipy}}




\bibliography{reference}{}
\bibliographystyle{aasjournal}



\end{document}